%% file: ddclust_spatial.tex
\documentclass[parskip=full,12pt]{article}

\usepackage{amssymb}
\usepackage{amsthm}
\usepackage{amsmath}
\usepackage{booktabs}
\usepackage{graphicx}
\usepackage{natbib}
\usepackage{url}
\usepackage{color,soul}
\usepackage{multirow}
\usepackage[top=1in, bottom=1.3in, left=1in, right=1in]{geometry}
\usepackage[linesnumbered,ruled]{algorithm2e}
\usepackage{tikz}
\usetikzlibrary{er,positioning,arrows.meta,calc}
\usepackage[labelformat=simple]{subcaption}

\newcommand{\E}{\text{E}}
\newcommand{\FA}{\text{FA}}
\newcommand{\prob}{\mathbf{P}}
\newcommand{\Vari}{\text{Var}}

\usetikzlibrary{er,positioning,arrows.meta,arrows,shapes,calc}

\tikzstyle{block} = [rectangle, draw, fill=white, 
    text width=7cm, text centered, rounded corners, minimum height=3em]

\tikzset{
     arrow/.style = { thick,  ->, >=Triangle},
}

\newcommand{\bfx}[1]{{\bf #1}}

\def\spacingset#1{\renewcommand{\baselinestretch}%
{#1}\small\normalsize} \spacingset{1}

\title{Applications of Clustering with Mixed Type Data in Life Insurance}

\author{Shuang Yin\thanks{Department of Statistics, University of Connecticut, 215 Glenbrook Road, Storrs, CT,  06269-4120, USA. Email: \texttt{shuang.yin@uconn.edu}.} \and Guojun Gan\thanks{Department of Mathematics, University of Connecticut, 341 Mansfield Road, Storrs, CT, 06269-1009, USA. Email: \texttt{guojun.gan@uconn.edu}.} \and Emiliano A. Valdez\thanks{Department of Mathematics, University of Connecticut, 341 Mansfield Road, Storrs, CT, 06269-1009, USA. Email: \texttt{emiliano.valdez@uconn.edu}.} \and Jeyaraj Vadiveloo\thanks{Department of Mathematics, University of Connecticut, 341 Mansfield Road, Storrs, CT, 06269-1009, USA. Email: \texttt{jeyaraj.vadiveloo@uconn.edu}.}}

\begin{document}

\maketitle

\begin{abstract}

Death benefits are generally the largest cash flow item that affects financial statements of life insurers where some still do not have a systematic process to track and monitor death claims experience. In this article, we explore data clustering to examine and understand how actual death claims differ from expected, an early stage of developing a monitoring system crucial for risk management. We extend the $k$-prototypes clustering algorithm to draw inference from a life insurance dataset using only the insured's characteristics and policy information without regard to known mortality. This clustering has the feature to efficiently handle categorical, numerical, and spatial attributes. Using gap statistics, the optimal clusters obtained from the algorithm are then used to compare actual to expected death claims experience of the life insurance portfolio. Our empirical data contains observations, during 2014, of approximately 1.14 million policies with a total insured amount of over 650 billion dollars. For this portfolio, the algorithm produced three natural clusters, with each cluster having a lower actual to expected death claims but with differing variability. The analytical results provide management a process to identify policyholders' attributes that dominate significant mortality deviations, and thereby enhance decision making for taking necessary actions.

\vspace{0.35cm}

\noindent \textbf{Keywords:} $k$-prototypes clustering; geospatial attributes, gap statistics; tracking and monitoring death claims.

\end{abstract}

\newpage

\section{Introduction and motivation} \label{sec:intro}

According to the Insurance Information Institute\footnote{\scriptsize{\url{https://www.iii.org/publications/2021-insurance-fact-book/life-health-financial-data/payouts}}}, the life insurance industry paid a total of nearly \$76 billion as death benefits in 2019. Life insurance is in the business of providing a benefit in the event of premature death, one that is understandably difficult to predict with certainty. Claims arising from mortality are not surprisingly the largest cash flow item that affects both the income statement and the balance sheet of a life insurer. Life insurance contracts are generally considered long duration where the promised benefit could be for an extended period of time before being realized. In effect, not only do life insurers pay out death claims in aggregate on a periodic basis, they are also obligated to have sufficient assets set aside as reserves to fulfill this long term obligation. See \citet{dickson2013}.

Every life insurer must have in place a systematic process of tracking and monitoring its death claims experience. This tracking and monitoring system is an important risk management tool. It should involve not only identifying statistically significant deviations of actual to expected experience, but also being able to understand and explain the effects of patterns. Such deviations might be considered normal patterns of deviation that are anomalies for short durations, while of more considerable importance are deviations considered to follow a trend for longer durations. 

Prior to sale, insurance companies exercise underwriting to identify the degree of mortality risk of applicants. As a consequence, there is a selection effect on the underlying mortality of life insurance policyholders; normally, the mortality of policyholders are considered better than the general population. However, this mortality selection wears off over time, and in spite of this selection, it is undeniably important for a life insurance company to have a monitoring system.  \cite{vadiveloo2014} listed some of these benefits and we reiterate their importance again as follows:
\begin{enumerate}
\itemsep0em
\item A tracking and monitoring system is a risk management tool that can assist insurers to take actions necessary to mitigate the economic impact of mortality deviations.
\item It is a tool for improved understanding of the emergence of death claims experience thereby helping an insurer in product design, underwriting, marketing, pricing, reserving, and financial planning.
\item It provides a proactive tool for dealing with regulators, credit analysts, investors, and rating agencies who may be interested in reasons for any volatility in earnings as a result of death claims fluctuations.
\item A better understanding of the company's emergence of death claims experience helps to improve its claims predictive models.
\item The results of a tracking and monitoring system provides the company a benchmark for its death claims experience that can be relatively compared with that of other companies in the industry.
\end{enumerate}

Despite these apparent benefits, several insurers still do not have a systematic process of tracking and monitoring death claims. Such a process clearly requires a meticulous investigation of historical death claims experience. In this article, we explore the use of data clustering to examine and understand how actual death claims differ from expected. By naturally subdividing the policyholders into clusters, this process of exploration through data clustering will provide us a better understanding of the characteristics of the life insurance portfolio according to their historical claims experience. This is an important early stage of developing a tracking and monitoring system that is a crucial part of risk management for a life insurer.

As information stored in data grows rapidly in the modern world, several industries, including the insurance industry, have started to implement practices to analyze datasets and to draw meaningful results for more effective decision making. The magnitude and scale of information from these datasets continue to increase at a rapid pace, and so does the ease of access. Data analytics have become an important function in every organization and how to deal with huge data sets has become an important issue. In many instances, information comes in unstructured forms so that unsupervised learning methods are instituted for preliminary investigation and examination. 

The most commonly used unsupervised learning technique is cluster analysis. It involves partitioning observations into groups or clusters where observations within each cluster are optimally similar while at the same time, observations between clusters are optimally dissimilar. Among many clustering algorithms developed in the past few decades, the $k$-means clustering algorithm (\cite{macqueen1967some}) is perhaps the simplest, most straightforward, and most popular method that efficiently partitions the data set into $k$ clusters. With $k$ initial centroids arbitrarily set, the $k$-means algorithm finds the locally optimal solutions by gradually minimizing the clustering error calculated according to numerical attributes. The technique has been applied in several disciplines including life insurance, e.g., \cite{thiprungsri2011cluster}, \cite{devale2012applications}, \cite{gan2013va}, and \cite{gan2016ed}. Despite its popularity, the algorithm has drawbacks that present challenges to our life insurance dataset: (i) it is particularly sensitive to the initial cluster assignment which is randomly picked, and (ii) it is unable to handle categorical attributes. While the $k$-prototypes clustering is lesser known, it provides the advantage of being able to handle mixed data types, including numerical and categorical attributes. For numerical attributes, the distance measure used may still be based on Euclidean. For categorical attributes, the distance measure used is based on the number of matching categories.

This paper extends the use of $k$-prototypes algorithm proposed by \cite{huang1997clustering} to provide insights and draw inference from a real-life dataset of death claims experience obtained from a portfolio of contracts of a life insurance company. The $k$-prototypes algorithm has been applied in marketing for segmenting customers to better understand product demands (\cite{hsu2007}) and in medical statistics for understanding hospital care practices (\cite{najjar2014}).  This algorithm integrates the procedures of $k$-means and $k$-modes to efficiently cluster datasets that contain, as earlier said, numerical and categorical variables; the nature of our data, however, contains a geospatial variable. The $k$-means can only handle numerical attributes while the $k$-modes can only handle categorical attributes. We therefore improve the $k$-prototypes clustering by adding a distance measurement to the cost function so that it can also deal with the geodetic distance between latitude-longitude
spatial data points. The latitude is a numerical measure of the distance of a location from far north or south of the equator; longitude is a numerical measure of the distance of a location from east-west of the ``meridians.'' Some work related to geospatial data clustering can be found in the Density-Based Spatial Clustering of Applications with Noise (DBSCAN) \citep{ester1996density} and in ontology \citep{wang2010geospatial}.

Our empirical data has been drawn from the life insurance portfolio of a major insurer and contains observations, during the third quarter of 2014, of approximately 1.14 million policies with a total insured amount of over 650 billion dollars. Using our empirical data, we applied the $k$-prototypes algorithm that ultimately yields to three optimal clusters determined using the concept of gap statistics. Shown to be an effective method for determining the optimal number of clusters, the gap statistic is based on evaluating ``the change in within-cluster dispersion with that expected under an appropriate reference null distribution'' \citep{tibshirani2001estimating}.

To provide further insights to the death claims experience of our life insurance data set, we compared the aggregated actual to expected deaths for each of the optimal clusters. For a life insurance contract, it is most sensible to measure the magnitude of deaths based on the face amount, and thus, we computed the ratio of the aggregated actual face amounts of those who died to the face amounts of expected deaths for each optimal cluster. Under some mild regularity conditions, necessary to prove normality, we are able to construct statistical confidence intervals of the ratio on each of the clusters thereby allowing us to draw inference as to the significant statistical deviations of the mortality experience for each of the optimal clusters. We provide details of the proofs for the asymptotic development of these confidence intervals in the Appendix. Each cluster showed different patterns of mortality deviation and we can deduce the dominant characteristics of the policies from this cluster-based analysis. The motivation is to assist the life insurance company gain some better understanding of potential favorable and unfavorable clusters.

For the rest of this paper, it has been organized as follows. In Section \ref{sec:data}, we briefly describe the real data set from an insurance company including the data elements and the preprocessing of the data in preparation for cluster analysis. In section \ref{sec:algo}, we provide details of the $k$-prototypes clustering algorithm and discuss how the balance weight parameter is estimated and how to choose the optimal number of clusters. In section \ref{sec:results}, we present the clustering results and discuss their implications and applications to monitoring the company's death claims experience. We conclude in Section \ref{sec:conclude}.

\section{Empirical data} \label{sec:data}

We illustrate  $k$-prototypes clustering algorithm based on the data set we obtained from an insurance company. This data set contains 1,137,857 life insurance policies issued in the third quarter of 2014. Each policy is described by 8 attributes with 5 categorical, 2 numerical data elements and longitude-latitude coordinates. Table \ref{tab:data summary} shows the description and basic summary statistics of each variable.  

\begin{table}[t]
  \centering
  \caption{Description of variables in the mortality dataset}
  \scalebox{0.87}{
    \begin{tabular}{rl|p{8.5em}cccr}
    \toprule
    \toprule
          & \multicolumn{1}{l}{\textbf{Categorical Variables}} & \multicolumn{1}{l}{\textbf{Description}} &       &       & \multicolumn{1}{l}{\textbf{Proportions}} &  \\
    \midrule
          &  \multirow{2}[0]{*}{\textbf{Gender}} & \multicolumn{1}{l}{\multirow{2}[0]{*}{Insured's sex}} &       & \multicolumn{1}{r}{Female} & 34.1\%  &  \\
          &       & \multicolumn{1}{l}{} &       & \multicolumn{1}{r}{Male} & 65.9\%  &  \\
          &       & \multicolumn{1}{r}{} &       &       &       &  \\
          & \multirow{2}[0]{*}{\textbf{Smoker Status }} & \multirow{2}[0]{*}{Insured's smoking status} &       & \multicolumn{1}{r}{Smoker                          } & 4.14\%   &  \\
          &       & \multicolumn{1}{l}{} &       & \multicolumn{1}{r}{Nonsmoker} & 95.86\%  &  \\
          &       & \multicolumn{1}{c}{} &       &       &       &  \\
          & \multirow{2}[0]{*}{\textbf{Underwriting Type }} & \multirow{2}[0]{*}{Type of underwriting requirement} &       & \multicolumn{1}{r}{Term conversion} & 4.52\%   &  \\
          &       & \multicolumn{1}{l}{} &       & \multicolumn{1}{r}{Underwritten} & 95.48\%  &  \\
          &       & \multicolumn{1}{l}{} &       &       &       &  \\
          & \multirow{2}[0]{*}{\textbf{Substandard Indicator }} & \multirow{2}[0]{*}{Indicator of substandard policies } &       & \multicolumn{1}{r}{Yes} & 7.76\%   &  \\
          &       & \multicolumn{1}{r}{} &       & \multicolumn{1}{r}{No } & 92.24\%  &  \\
          &       & \multicolumn{1}{r}{} &       &       &       &  \\
          & \multirow{1}[1]{*}{\textbf{Plan}} & \multicolumn{1}{l}{\multirow{1}[1]{*}{Plan type}} &       & \multicolumn{1}{r}{Term} & 74.28\%  &  \\
          &       & \multicolumn{1}{l}{} &       & \multicolumn{1}{r}{ULS} & 14.55\%  &  \\
          &       & \multicolumn{1}{l}{} &       & \multicolumn{1}{r}{VLS} & 11.17\%  &  \\
    \midrule
    \midrule
          & \multicolumn{1}{l}{\textbf{Continuous Variables}} & \multicolumn{1}{r}{} & \textbf{Minimum} & \textbf{Mean} & \textbf{Maximum} &  \\
    \midrule
          & \textbf{Issue Age} & Policyholder's age at issue & 0     & 43.62  & 90    &  \\
          &       & \multicolumn{1}{r}{} &       &       &       &  \\
          & \textbf{Face Amount} & Amount of sum insured at issue & 215   & 529,636 & 100,000,000 &  \\
    \bottomrule
    \bottomrule
    \end{tabular}%
}
  \label{tab:data summary}
\end{table}%

\begin{figure}[!ht]
    \centering
    \includegraphics[scale=0.65]{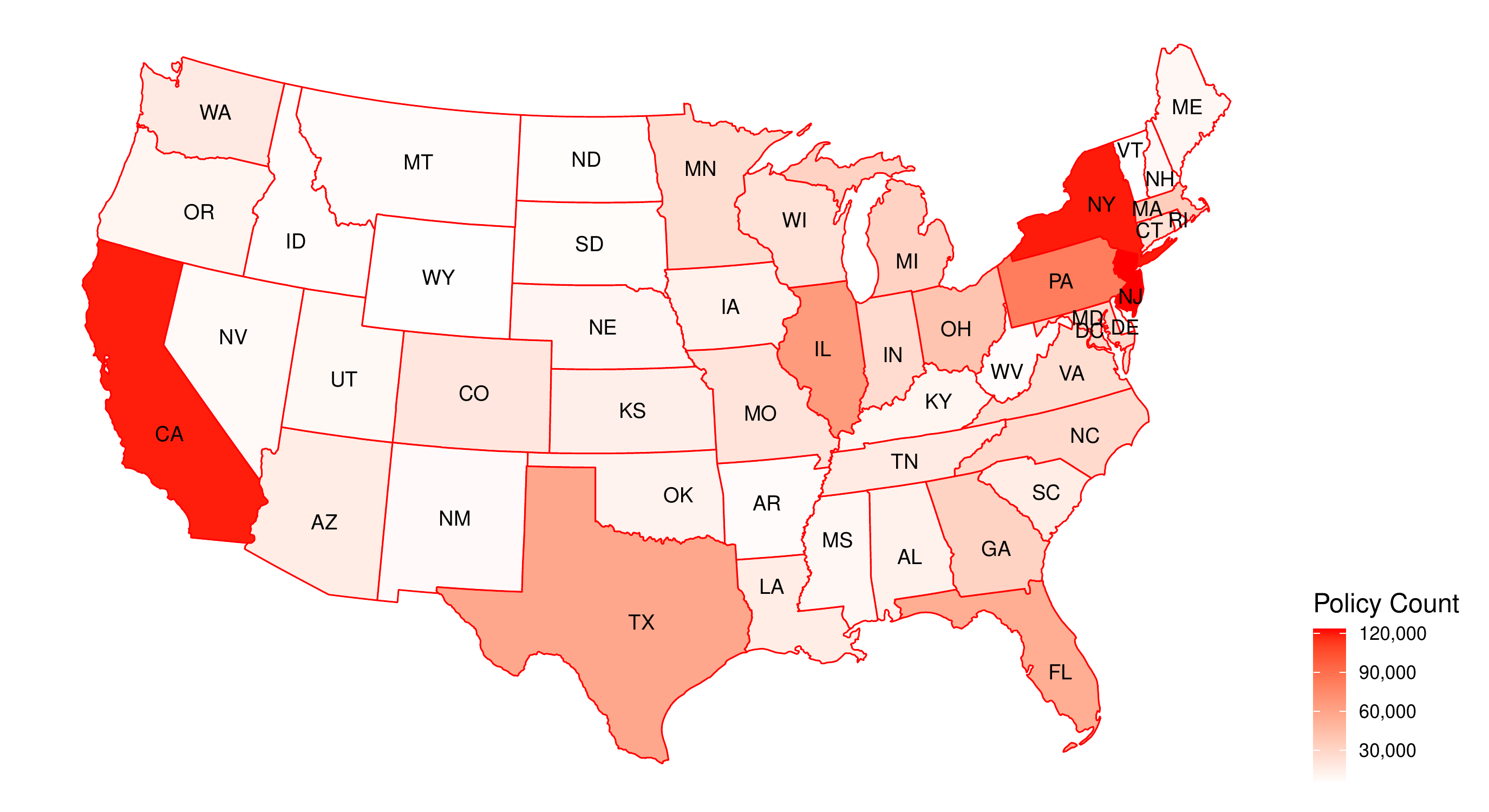}
    \caption{U.S Heatmap of Policy Frequency}
    \label{fig:state_heatmap}
\end{figure}

Figure \ref{fig:state_heatmap} provides a visualization of the distribution of the policies across the states. We only kept the policies issued in the continental United States, and therefore, excluded the policies issued in Alaska, Hawaii, and Guam. First, the frequency of policies observed from these states are not materially plentiful. Second, since these states or territories are outside the mainland United States, geodetic measurements are distorted and clustering results may become less meaningful. The saturated color indicates a high frequency of the policy distributed in a particular state. The distribution of the policy count is highly skewed, with New York, New Jersey, California, and Pennsylvania having insureds significantly more than other states. The spatial attributes are represented by latitude and longitude coordinate pairs.

The Insured's sex indicator \textit{Gender} is also a discrete variable with 2 levels, Female and Male, with the number of males almost twice as many as females. \textit{Smoker Status} indicates the insured's smoking status with $95.86\%$ nonsmokers and the remaining $4.14\%$ smokers. The variable \textit{Underwriting Type} reflects two types of underwriting: $95.48\%$ of the policies were fully underwritten at issue while the remaining $4.52\%$ are term conversions. Term conversions refer to those policies originally with a fixed maturity (or term) but were converted into permanent policies at a later date from issue, without any additional underwriting. The variable \textit{Substandard Indicator} indicates whether policy has been issued as substandard or not. Substandard policies are issued after an underwriting is performed that have expected mortality worse than standard policies. Substandard policies come with an extra premium. In our dataset, there are about $7.76\%$ policies considered substandard and the remaining $92.24\%$ are standard. The variable \textit{Plan} has three levels: Term Insurance Plan (Term), Universal Life with Secondary Guarantees (ULS) and Variable Life with Secondary Guarantees (VLS).

In our dataset, there are two continuous variables. The variable \textit{Issue Age} refers to the policyholder's age at the time of issue; the range of issue ages is from as young as a newborn to as old as 90 years, with an average of about 44 years old. The variable \textit{Face Amount} refers to the amount of sum insured either fixed at policy issue or accumulated to this level at the most recent time of valuation. As is common with data clustering, we standardized these two continuous variables by rescaling the values in order to be in the range of $[0,1]$. The general formula used in our normalization is 
\[
x_{new} = \frac{x-\text{min}(x)}{\text{max}(x)-\text{min}(x)}, 
\]
where $x$ is the original value and $x_{new}$ is the standardized (or normalized) value. However, for the variable \text{Face Amount}, we find few extreme values that may be further distorting the spread or range of possible values. To fix this additional concern, we take the logarithm of the original values before applying the normalization formula:
\[
x_{new} = \frac{\log(x)-\text{min}(\log(x))}{\text{max}(\log(x))-\text{min}(\log(x))}.
\]

\section{Data clustering algorithms} \label{sec:algo}

Data clustering refers to the process of dividing a set of objects into homogeneous groups or clusters \citep{gan2007, gan2011} using some similarity criterion. Objects in the same cluster are more similar to each other than to objects from other clusters. Data clustering is an unsupervised learning process and is often used as a preliminary step for data analytics. In bioinformatics, for example, data clustering is used to identify the patterns hidden in gene expression data \citep{maccuish2010}. In big data analytics, data clustering is used to produce a good quality of clusters or summaries for big data to address the storage and analytical issues \citep{fahad2014big}. In actuarial science, data clustering is also used to select representative insurance policies from a large pool of policies in order to build predictive models \citep{gan2013va, gan2015ns, gan2016ed}.

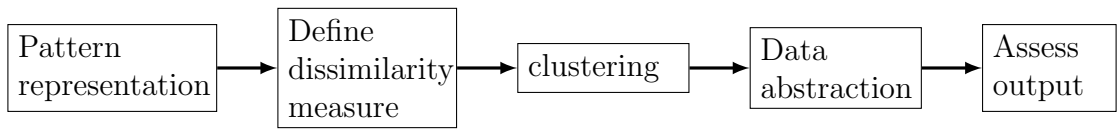
\begin{figure}[htbp]
\centering
\begin{tikzpicture}[node distance=0.8cm]
\node[rectangle, draw=black, text width=2.5cm](s2){Pattern\\ representation};
\node[right=of s2, rectangle, draw=black, text width=2.1cm](s3){Define\\ dissimilarity\\ measure };
\node[right =of s3, rectangle, draw=black, text width=2cm](s4){clustering};
\node[right =of s4, rectangle, draw=black, text width=2cm](s5){Data\\ abstraction};
\node[right =of s5, rectangle, draw=black, text width=1.5cm](s6){Assess\\ output};
\draw[->,>=latex,very thick] (s2)--(s3);
\draw[->,>=latex,very thick] (s3)--(s4);
\draw[->,>=latex,very thick] (s4)--(s5);
\draw[->,>=latex,very thick] (s5)--(s6);
\end{tikzpicture}
\caption{A typical data clustering process.}\label{fig:process}
\end{figure}

Figure \ref{fig:process} shows a typical clustering process described in \cite{jain1999review}. The clustering process consists of five major steps: pattern representation, dissimilarity measure definition, clustering, data abstraction, and output assessment. In the pattern representation step, the task is to determine the number and type of the attributes of the objects to be clustered. In this step, we may extract, select, and transform features to identify the most effective subset of the original attributes to use in clustering. In the dissimilarity measure definition step, we select a distance measure that is appropriate to the data domain. In the clustering step, we apply a clustering algorithm to divide the data into a number of meaningful clusters. In the data abstraction step, we extract one or more prototypes from each cluster to help comprehend the clustering results. In the final step, we use some criteria to assess the clustering results.

Clustering algorithms can be divided into two categories: partitional and hierarchical clustering algorithms. A partitional clustering algorithm divides a dataset into a single partition; while a hierarchical clustering algorithm divides a dataset into a sequence of nested partitions. In general, partitional algorithms are more efficient than hierarchical algorithms because the latter usually require calculating the pairwise distances between all the data points.

\subsection{The $k$-prototypes algorithm}

The $k$-prototypes algorithm \citep{huang1998extensions} is an extension of the well-known $k$-means algorithm for clustering mixed type data. In the $k$-prototypes algorithm, the prototype is the center of a cluster, just as the mean is the center of a cluster in the $k$-means algorithm.

To describe the $k$-prototypes algorithm, let $\{X_{i j}\}$, $i=1, 2, \dots, n, j=1, 2, \dots,d$ denote a dataset containing $n$ observations. Each observation is described by $d$ variables, including $d_1$ numerical variables, $d_2-d_1$ categorical variables, and $d-d_2=2$ spatial variables. Without loss of generality, we assume that the first $d_1$ variables are numerical, the remaining $d_2-d_1$ variables are categorical, and the last two variables are spatial. Then the dissimilarity measure between two points $\bfx{x}$ and $\bfx{y}$ used by the $k$-prototypes algorithm is defined as follows:
\begin{equation}\label{eq:dist}
D(\bfx{x}, \bfx{y}) = \sum_{j=1}^{d_1} (x_j - y_j)^2 + \lambda_1 \sum_{j=d_1+1}^{d_2} \delta_1(x_j,y_j)+\lambda_2  \delta_2(x^*, y^*),
\end{equation}
where $\lambda_1$ and $\lambda_2$ are balancing weights with respect to numerical attributes that is used to avoid favoring types of variables other than numerical, $\delta_1(\cdot,\cdot)$ is the simple-matching distance defined as
\[
\delta_1(x_j,y_j) = \left\{
\begin{array}{ll}
1, & \hbox{if $x_j\ne y_j$,}\\
0, & \hbox{if $x_j=y_j$.}
\end{array}
\right.
\]
and $\delta_2(\cdot,\cdot)$ returns the spatial distance between two points with latitude-longitude coordinates using Great Circle distance (WGS84 ellipsoid) methods. We have $x^*=(x_{d_2+1}, x_{d_2+2})$, $y^*=(y_{d_2+1}, y_{d_2+2})$ and radius of the Earth $r= 6378137$m from WGS84 axis \citep{carter2002great}, 
\begin{align*}
    \Delta a & = x_{d_2+2} - x_{d_2+1} \\
    \Delta b & = y_{d_2+2}-y_{d_2+1} \\
    A & = \cos(x_{d_2+2}) \sin (\Delta b) \\
    B &= \sin(\Delta a) +\cos(x_{d_2+2}) \sin (x_{d_2+1})[1-\cos(\Delta b)] \\
    \Phi_{2, 1} & = \tan^{-1} (A/B) \\
    \Theta_{2,1} &= \tan^{-1} \left [ \frac{B \cos(\Phi_{2,1})+A\sin(\Phi_{2,1})}{\cos(\Delta A)-\cos(x_{d_2+1})\cos(x_{d_2+2})[1-\cos(\Delta a)]} \right ] \\
    \delta_2(x^*, y^*) & =r(1-f) \times \Theta_{2,1}
\end{align*}
where $f$ is the flattening of the Earth (use $1/298.257223563$ according to WGS84). WGS84 is the common system of reference coordinate used by the Global Positioning System (GPS), and is also the standard set by the U.S. Department of Defense for a global reference system for geospatial information.

The $k$-prototypes algorithm aims to minimize the following objective (cost) function:
\begin{equation}\label{eq:obj}
P(U,Z) = \sum_{i=1}^n \sum_{l=1}^k u_{il} D(\bfx{x}_i, \bfx{z}_l),
\end{equation}
where $U=(u_{il})_{i=1:n, l=1:k}$ is an $n\times k$ partition matrix, $Z=\{\bfx{z}_1,\bfx{z}_2,\ldots,\bfx{z}_k\}$ is a set of prototypes, and $k$ is the desired number of clusters. The $k$-prototypes algorithm employs an iterative process to minimize this objective function. The algorithm starts with $k$ initial prototypes selected randomly from the dataset. Given the set of prototypes $Z$, the algorithm then updates the partition matrix as follows:
\begin{equation}\label{eq:U}
u_{il} = \left\{
\begin{array}{ll}
1, & \hbox{if $D(\bfx{x}_i, \bfx{z}_l) = \min_{1\le s\le k} D(\bfx{x}_i, \bfx{z}_s)$,}\\
0, & \hbox{if otherwise.}
\end{array}
\right.
\end{equation}
Given the partition matrix $U$, the algorithm updates the prototypes as follows:
\begin{subequations}\label{eq:Z}
\begin{equation}\label{eq:Zn}
z_{lj} = \frac{\sum_{i=1}^n u_{il} x_{ij} }{\sum_{i=1}^n u_{il}},\quad 1\le j\le d_1,
\end{equation}
\begin{equation}\label{eq:Zc}
z_{lj} = \operatorname{mode } \{x_{ij}: u_{il}=1 \},\quad d_1+1\le j\le d_2,
\end{equation}
\begin{equation}\label{eq:Zs}
(z_{l, d_2+1}, z_{l, d_2+2}) =  \{(x_{i,d_2+1}, x_{i,d_2+2}) \bigr| \min(\delta_{2}(\bfx{x}^*, z_l^*)) \}, \end{equation}
\end{subequations}
where $\bfx{x}^*=\{(x_{1, d_2+1},x_{1, d_2+2}), (x_{2, d_2+1},x_{2, d_2+2}), \dots, (x_{n, d_2+1},x_{n, d_2+2})\}$ and $z_l^*=(z_{l, d_2+1},z_{l, d_2+2})$. When $\delta_2$ is calculated, we exclude the previous spatial prototype. The numerical components of the prototype of a cluster are updated to the means, the categorical components are updated to the modes, and the new spatial prototype is the coordinate closest to the previous one.

Algorithm \ref{alg:kprototype} shows the pseudo-code of the $k$-prototypes algorithm. A major advantage of the $k$-prototypes algorithm is that it is easy to implement and efficient for large datasets. A drawback of the algorithm is that it is sensitive to the initial prototypes, especially when $k$ is large.

\begin{algorithm}[H]
\KwIn{A dataset $X$, $k$}
\KwOut{$k$ clusters}
Initialize $\bfx{z}_1,\bfx{z}_2,\ldots,\bfx{z}_k$ by randomly selecting $k$ points from $X$\;
\Repeat{No further changes of cluster membership}{
Calculate the distance between $\bfx{x}_i$ and $\bfx{z}_j$ for all $1\le i\le n$ and $1\le j\le k$\;
Update the partition matrix $U$ according to Equation \eqref{eq:U}\;
Update cluster centers $Z$ according to Equation \eqref{eq:Z}\;
}
Return the partition matrix $U$ and the cluster centers $Z$\;
\caption{Pseudo-code of the $k$-prototypes algorithm.}\label{alg:kprototype}
\end{algorithm}

\subsection{Determining the parameters $\lambda_1$ and $\lambda_2$}
The cost function in Equation (\ref{eq:obj}) can be further rewritten as: 
   \begin{equation*}
    \begin{aligned}
    P(U,Z)  &= \sum_{l=1}^k  \sum_{i=1}^n u_{il}\left \{ \sum_{j=1}^{d_1} (x_{ij}-z_{lj})^2+
            \lambda_1 \sum_{j=d_1+1}^{d_2} \delta_1(x_{ij},z_{lj})+\lambda_2  \delta_2(x_i^*, z_l^*)
            \right \},
    \end{aligned}
\end{equation*}
where $x_i^*=(x_{i, d_2+1}, x_{i, d_2+2})$ and the inner term 
\begin{equation*}
    \begin{aligned}
    D_l 
    &= \sum_{i=1}^n u_{il}\left \{ \sum_{j=1}^{d_1} (x_{ij}-z_{lj})^2+
    \lambda_1 \sum_{j=d_1+1}^{d_2} \delta_1(x_{ij},z_{lj})+\lambda_2 \, \delta_2(x_i^*, z_l^*)\right \} \\
    &=D_l^n + D_l^c +D_l^s  \\
    \end{aligned}
\end{equation*}
is the total cost when $X$ is assigned to cluster $l$. Note that we can subdivide these measurements into
\begin{align*}
D_l^n &= \sum_{i=1}^n u_{il} \sum_{j=1}^{d_1} (x_{ij}-z_{lj})^2, \\
D^c &= \sum_{i=1}^n u_{il} \, \lambda_1 \sum_{j=d_1+1}^{d_2} \delta_1(x_{ij},z_{lj}), \text{ and} \\
D_l^s &= \sum_{i=1}^n u_{il} \, \lambda_2 \, \delta_2(x_i^*, z_l^*),
\end{align*}
that represent the total cost from the numerical, categorical, and spatial attributes, respectively. 

It is easy to show that the total cost $D_l$ is minimized by individually minimizing $D_l^n$, $D_l^c$, and $D_l^s$ (\cite{huang1997clustering}). $D_l^n$ can be minimized through Equations \eqref{eq:Zn}. $D_l^c$, the total cost from categorical attributes of $X$, can be rewritten as
\begin{align*}
    D_l^c &= \lambda_1 \sum_{i=1}^n  u_{il} \sum_{j=d_1+1}^{d_2} \delta_1(x_{ij},z_{lj}) \\
    &= \lambda_1
    \sum_{i=1}^n \sum_{j=d_1+1}^{d_2}  \{ 1\cdot (1-\prob(x_{ij} = z_{lj}|l)) + 0\cdot  \prob(x_{ij} =z_{lj}|l) \} \\
    &= \lambda_1 \sum_{i=1}^n \sum_{j=d_1+1}^{d_2} \{1-\prob(x_{ij} =z_{lj}|l)\}\\
    &= \lambda_1 \sum_{j=d_1+1}^{d_2} n_l
    \{1-\prob(z_{lj} \in A_j |l)\},
\end{align*}
where $A_j$ is the set of all unique levels of the $j$th categorical attribute of $X$ and  $\prob(z_{lj} \in A_j|l)$ denotes the probability that the $j$th categorical attribute of prototype $\bfx{z}_l$ occurs given cluster $l$.  $\lambda_1$ and $\lambda_2$ are chosen to prevent over-emphasizing either categorical or spatial with respect to numerical attributes and hereby are dependent on the distributions of those numerical attributes (\cite{huang1997clustering}). In the R package, \cite{clustMixType} suggested the value of $\lambda_1$ as the ratio of average of variance of numerical variables to the average concentration of categorical variables:
\begin{equation*}
    \hat{\lambda}_1  = \frac{\frac{1}{d_1}\sum_{j=1}^{d_1} \Vari(\mathbf{x}_j)}{\frac{1}{d_2-(d_1+1)}\sum_{j=d_1+1}^{d_2}\sum_k q_{jk} (1-q_{jk})} \\
      = \frac{\frac{1}{d_1}\sum_{j=1}^{d_1} \Vari(\mathbf{x}_j)}{\frac{1}{d_2-(d_1+1)}\sum_{j=d_1+1}^{d_2}(1-\sum_k q_{jk}^2)}, 
\end{equation*}
where $q_{jk}$ is the frequency of the $k$th level of the $j$th categorical variable. See also \cite{szepannek2018}. For each categorical variable, we consider it to have a distribution with a probability of each level to be the frequency of this level. For example, the categorical data element \textit{Plan} has three levels: Term, Universal life with secondary guarantees (ULS) and Variable life with secondary guarantees (VLS). Then the concentration of \textit{Plan} can be measured by Gini impurity: $\sum_{k=1}^3 q_{jk}(1-q_{jk})=1-\sum_{k=1}^3 q^2_{jk}$. Therefore, under the condition that all the variables are independent, the total Gini impurity  for categorical variables is $\sum_{j=d_1+1}^{d}(1-\sum_k q_{jk}^2)$, since $\sum_{k=1}^3 q_{jk} = 1$. The average of the total variance for the numerical variables $\frac{1}{d_1}\sum_{j=1}^{d_1} \Vari(\mathbf{x}_j)$ can be considered to be the estimate of the population variance. Subsequently, $\hat{\lambda}_1$ becomes a reasonable estimate and is easy to calculate.

Similarly, $\displaystyle \hat{\lambda}_2 = \frac{\frac{1}{d_1}\sum_{j=1}^{d_1} \Vari(\mathbf{x}_j)}{ \Vari(\delta_2(\bfx{x}^*, \text{center}))}$, where the concentration of spatial attributes is estimated by the variance of the Great Circle distances between $\bfx{x}^*$ and the center of the total longitude-latitude coordinates.

\subsection{Determining the optimal number of clusters}

\begin{figure}[htbp]
    \centering
    \begin{subfigure}[b]{0.49\textwidth}
    \includegraphics[width=\textwidth]{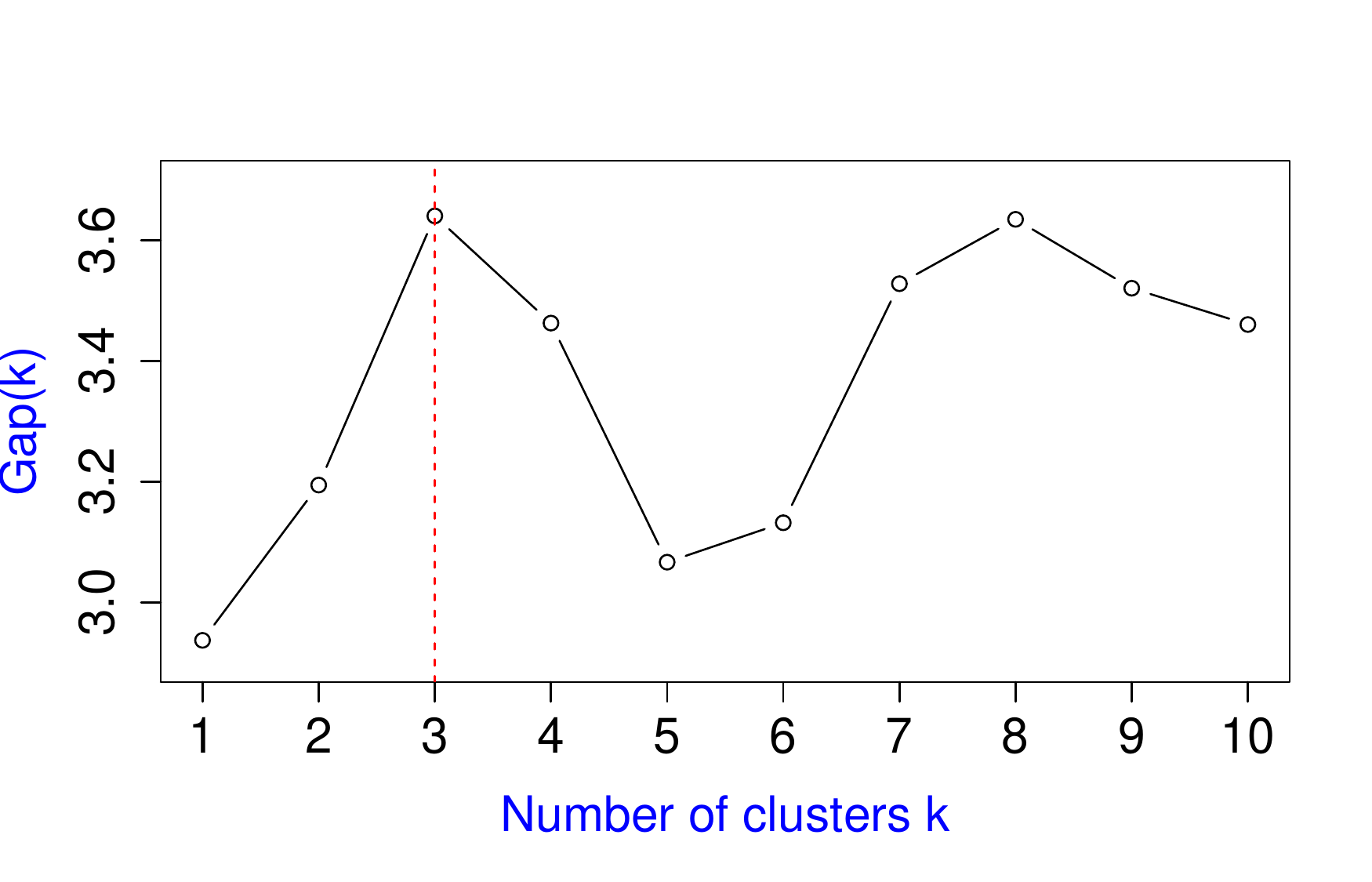}
    \caption{}
    \end{subfigure}
    \begin{subfigure}[b]{0.49\textwidth}
      \includegraphics[width=\textwidth]{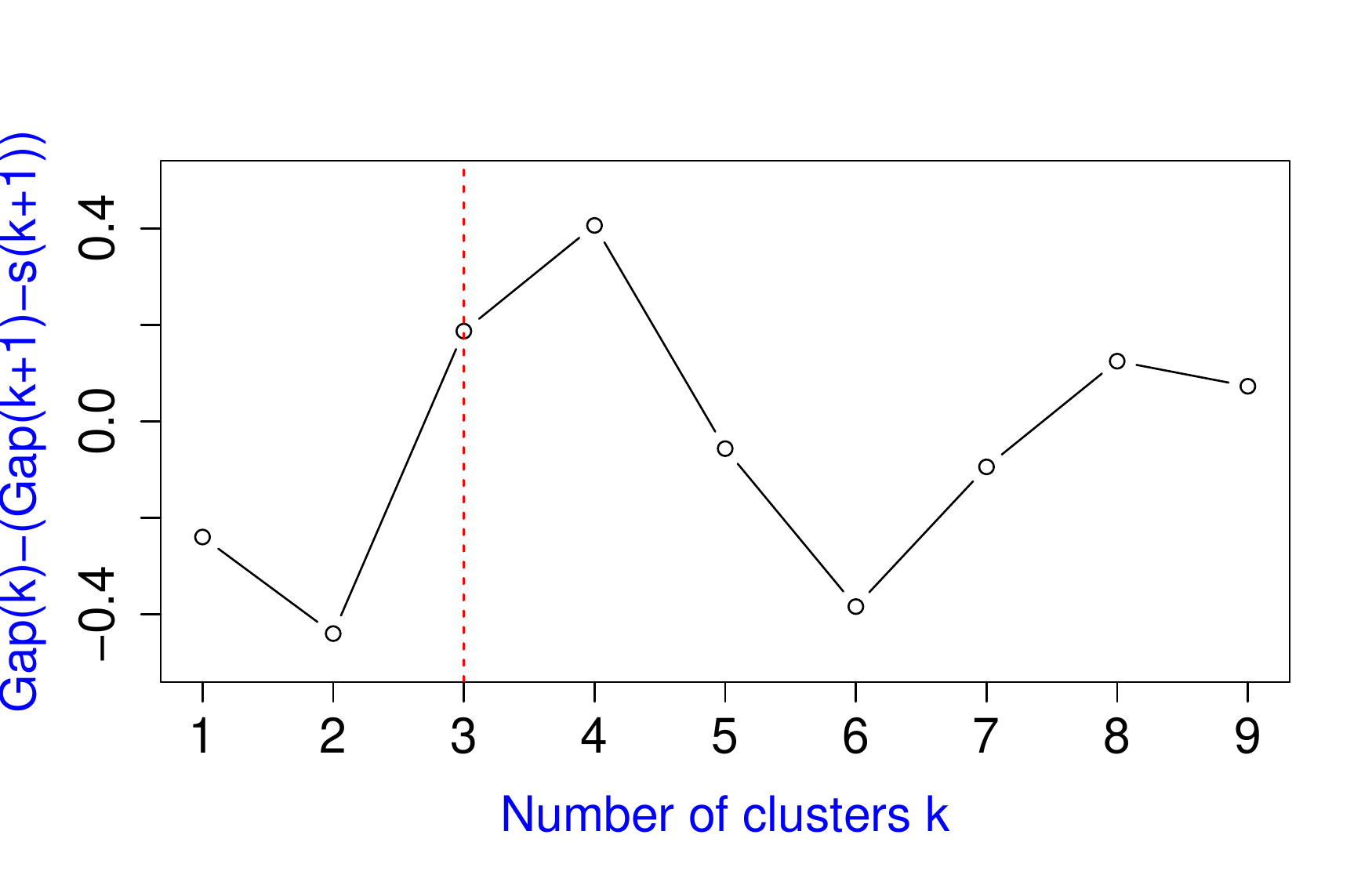}
      \caption{}
      \end{subfigure}
      \caption{(a) Gap statistics in terms of the corresponding number of clusters and (b) Results of choosing optimal number of clusters.}
      \label{fig:gap_k}
\end{figure}

As alluded in Section \ref{sec:intro}, the gap statistic is used to determine the optimal number of clusters. Data $\bfx{X}=\{X_{i j}\}$, $i=1, 2, \dots, n, j=1, 2, \dots,d$ consists of $d$ features measured on $n$ independent observations. $D_{i j}$ denotes the distance, defined in Equation \ref{eq:dist}, between observation $i$ and $j$. Suppose that we have partitioned the data into $k$ clusters $C_1, \dots, C_k$ and $n_l=|C_l|$. Let
\begin{equation*}
    D^w_l= \sum_{i, j \in C_l} D_{i j}
\end{equation*}
be the sum of the pairwise distance for all points within cluster $l$ and set 
\begin{equation*}
    W_k(\bfx{X}) = \sum_{l=1}^k \frac{1}{2n_l} D^w_l.
\end{equation*}
The idea of the approach is to standardized the comparison of $\log(W_k)$ with its expectation under an appropriate null reference distribution of the data. We define
\begin{equation*}
    \text{Gap}(k) = \E[\log(W_k(\bfx{X}^*))] -\log(W_k(\bfx{X})),
\end{equation*}
where $\E[\log(W_k(\bfx{X}^*))]$ denotes the average $\log(W_k)$ of the samples $\bfx{X}^*$ generated from the reference distribution with predefined $k$. The gap statistic can be calculated by the following steps:
\begin{itemize}
    \item Set $k=1, 2, \dots, 10$;
    \item Run $k$-prototypes algorithm and calculate $\log(W_k)$ under each $k=1, 2, \dots, 10$ for the original data $\bfx{X}$; 
    \item For each $b=1, 2, \dots, B$, generate a reference data set $\bfx{X}_b^*$ with sample size $n$. Run the clustering algorithm under the candidate $k$ values and compute 
$$
\E[\log(W_k(\bfx{X}^*))]=\frac{1}{B}\sum_{b=1}^B \log(W_k(\bfx{X}_b^*))
$$
and $\text{Gap}(k)$;
    \item  Define $s(k)=\left(\sqrt{1+1/B}\right)\times \text{sd}(k)$, where \\
    $\text{sd}(k)=\sqrt{(1/B) \sum_{b=1}^B (\log (W_k(\bfx{X}_b^*)) - \E[\log(W_k(\bfx{X}^*))])^2}$; and
    \item Choose the optimal number of clusters as the smallest $k$ such that $\text{Gap}(k) \ge \text{Gap}(k+1)-s(k+1)$.
\end{itemize}
This estimate is broadly applicable to any clustering method and distance measure $D_{i j}$. We use $B=50$ and randomly draw $10\%$ of the data set using stratified sampling to keep the same proportion of each attribute. The Gap and the quantity $\text{Gap}(k)-(\text{Gap}(k+1)-s(k+1))$ against the number of clusters $k$ are shown in Figure \ref{fig:gap_k}. The Gap statistic clearly peaks at $k=3$ and the criteria for choosing $k$ displayed in the right panel. The correct $k=3$ is the smallest for which the quantity  $\text{Gap}(k)-(\text{Gap}(k+1)-s(k+1))$ becomes positive. 

There is the possible drawback of the highly sensitivity of the initial choice of prototypes. In order to minimize the impact, we run the $k$-prototypes algorithm with correct $k=3$ starting with 20 different initializations and then choose the one with the smallest total sum squared errors.  

\section{Implications and applications of numerical results} \label{sec:results}

\subsection{Clustering results}

Using our mortality dataset with eight different attributes that are mixed type (numerical, categorical and spatial), we concluded as detailed in the previous section that three clusters are formed. Table \ref{tab:clustersize} displays the size  and membership degree of each cluster. Cluster 3 has the largest membership of nearly 57\% of the total observations, while Clusters 1 and 2 are partitioned with 30.1\% and 13.0\% memberships, respectively.

\begin{table}[htbp]
  \centering
  \caption{Size and percentage for each of the three optimal clusters}
    \begin{tabular}{lrrr}
          & \multicolumn{1}{l}{Cluster 1} & \multicolumn{1}{l}{Cluster 2} & \multicolumn{1}{l}{Cluster 3}  \\
    \midrule
    \midrule
    number of observations  & 342,518 & 147,561 & 647,778  \\
    percentage  & 30.10\% & 12.97\% & 56.93\% \\
    \bottomrule
    \bottomrule
    \end{tabular}%
  \label{tab:clustersize}%
\end{table}%

\begin{figure}[htbp]
    \centering
    \includegraphics[width=1.0\textwidth]{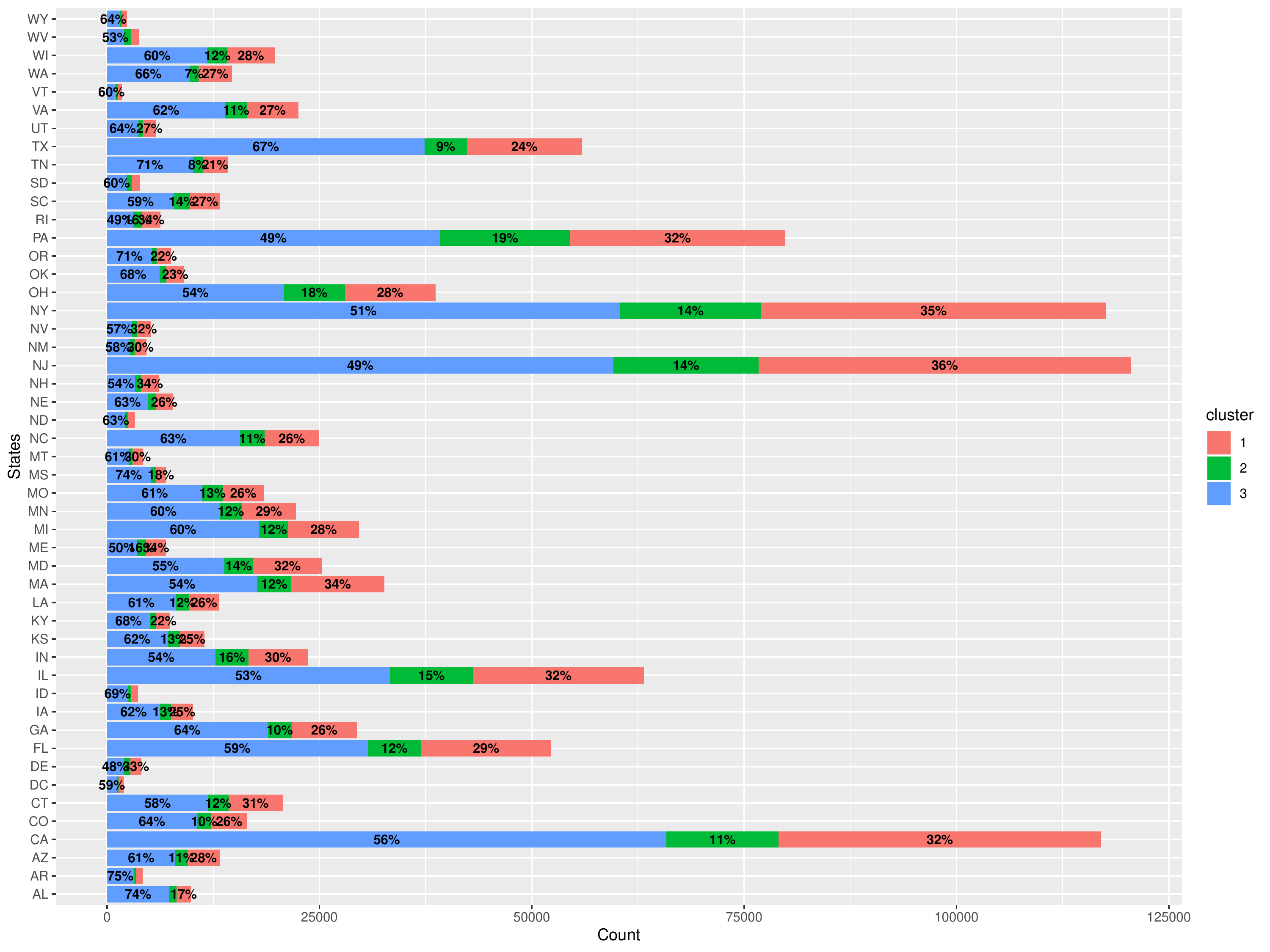}
    \caption{Distribution of the variable \textit{States} in each of the optimal clusters}
    \label{fig:cluster_states}
\end{figure}

\begin{figure}[htbp]
\centering
\begin{subfigure}[c]{0.3\linewidth}
 \includegraphics[width=1\textwidth]{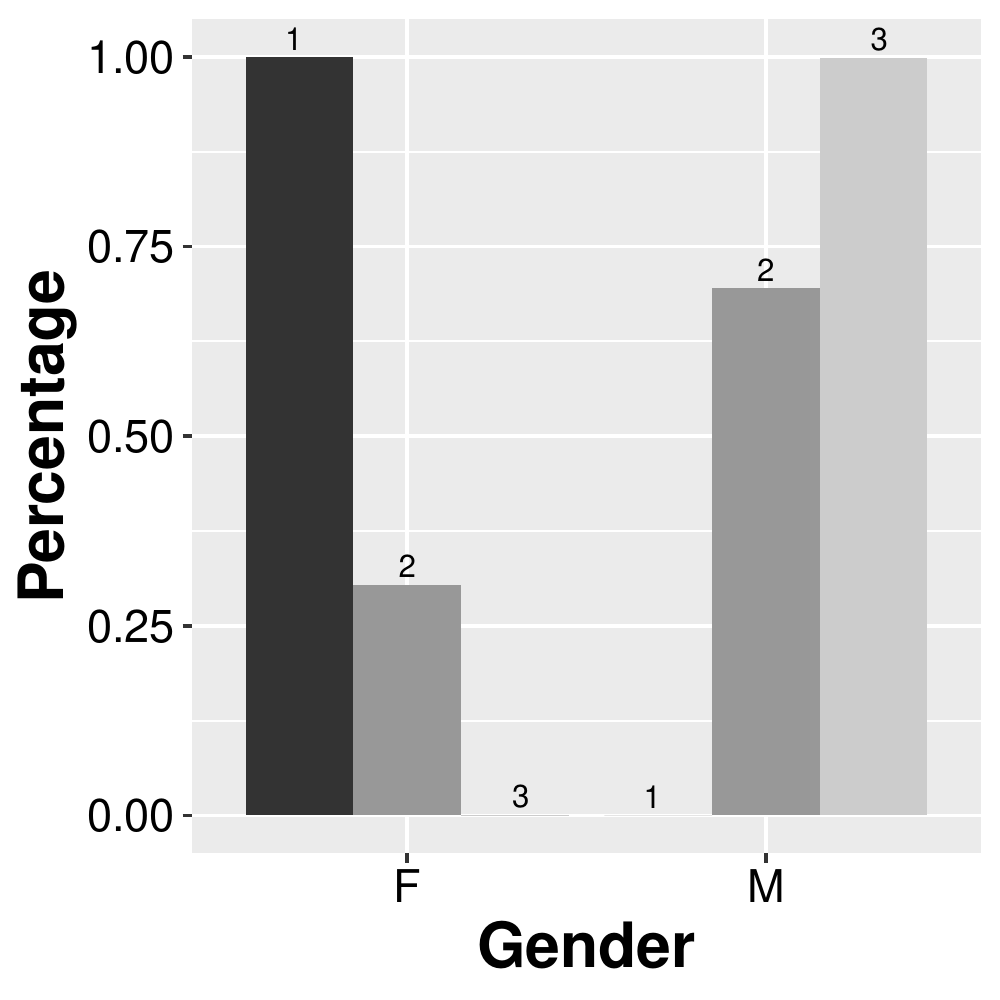}
 \label{fig: gender}
\end{subfigure}
\hfill
\begin{subfigure}[l]{0.3\linewidth}
 \includegraphics[width=1\textwidth]{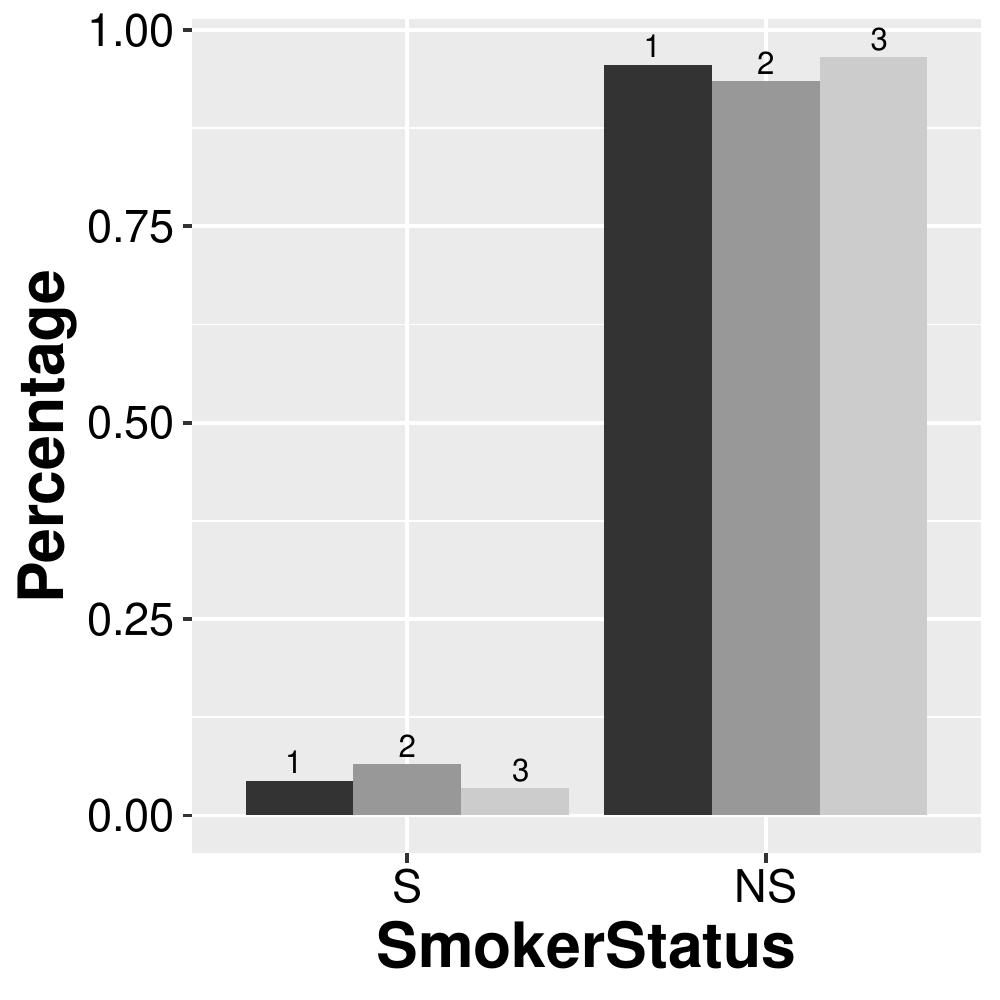}
 \label{fig: smoker}
\end{subfigure}%
\hfill
\begin{subfigure}[c]{0.3\linewidth}
 \includegraphics[width=1\textwidth]{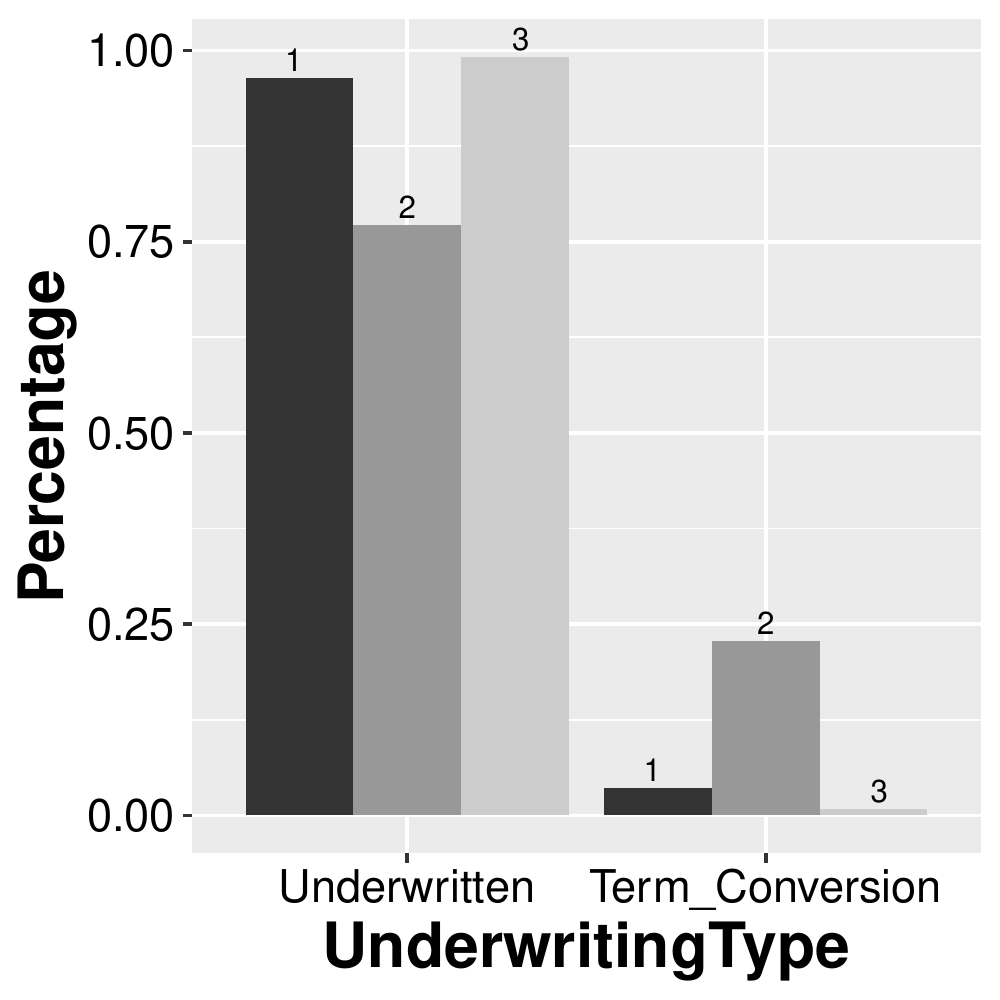}
 \label{fig: underwriting}
\end{subfigure}%
\vskip 1.25em
\begin{subfigure}[l]{0.3\linewidth}
 \includegraphics[width=1\textwidth]{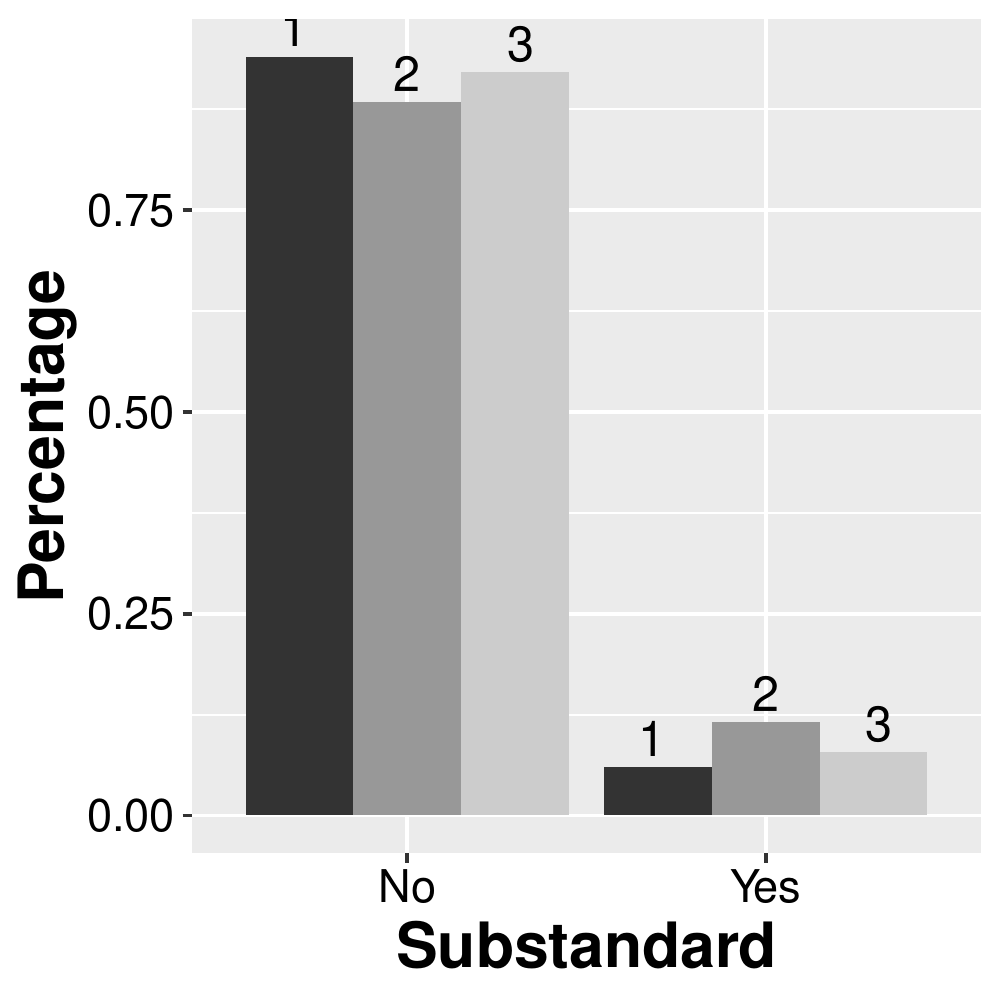}
 \label{fig: substandar}
\end{subfigure}%
\hfill
\begin{subfigure}[c]{0.3\linewidth}
 \includegraphics[width=1\textwidth]{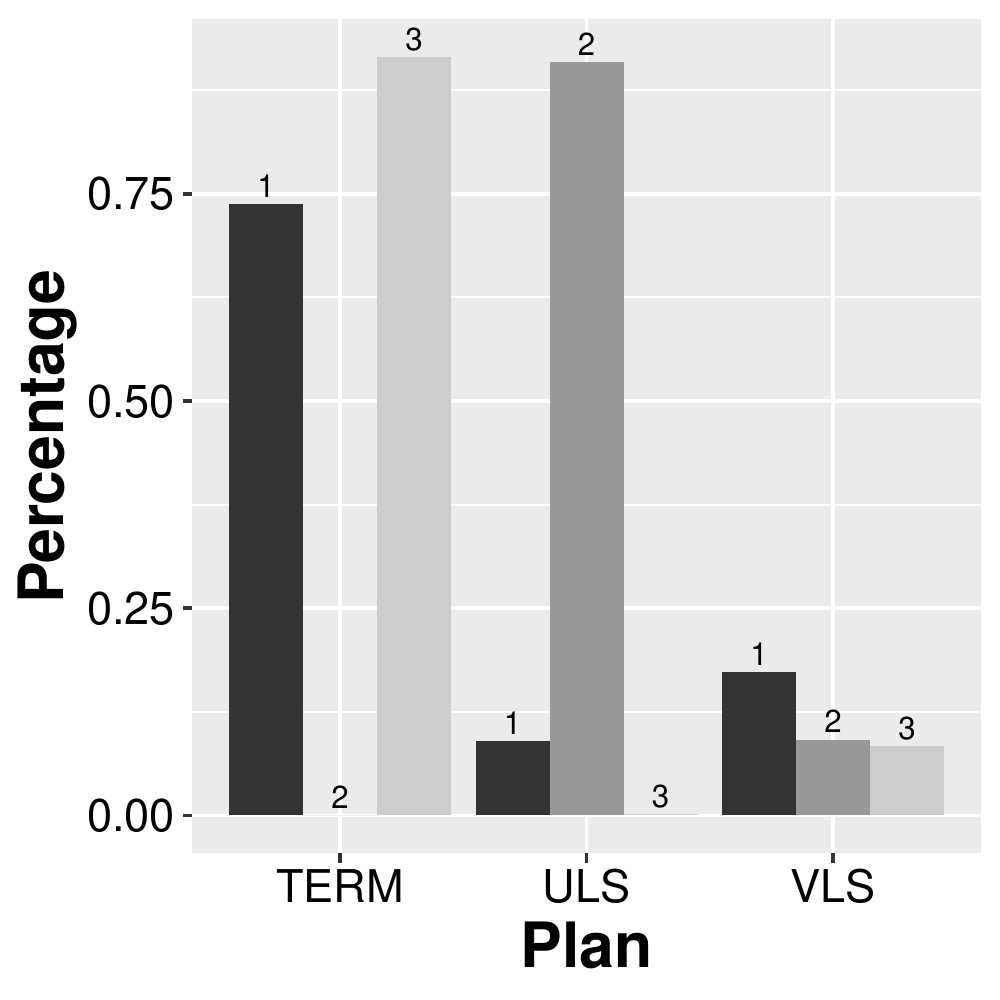}
 \label{fig: plan}
\end{subfigure}%
\hfill
\begin{subfigure}[l]{0.3\linewidth}
 \includegraphics[width=1\textwidth]{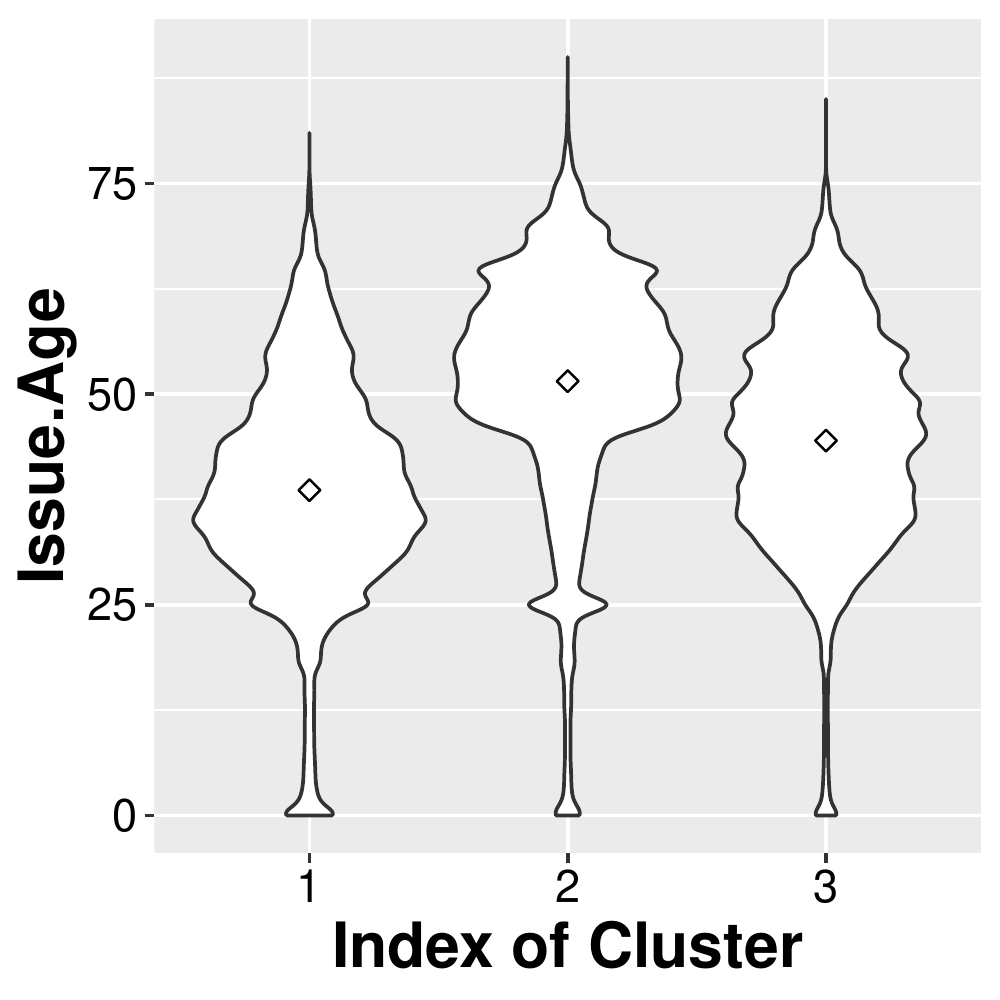}
 \label{fig: age}
\end{subfigure}%
\hfill
\begin{subfigure}[c]{0.3\linewidth}
 \includegraphics[width=1\textwidth]{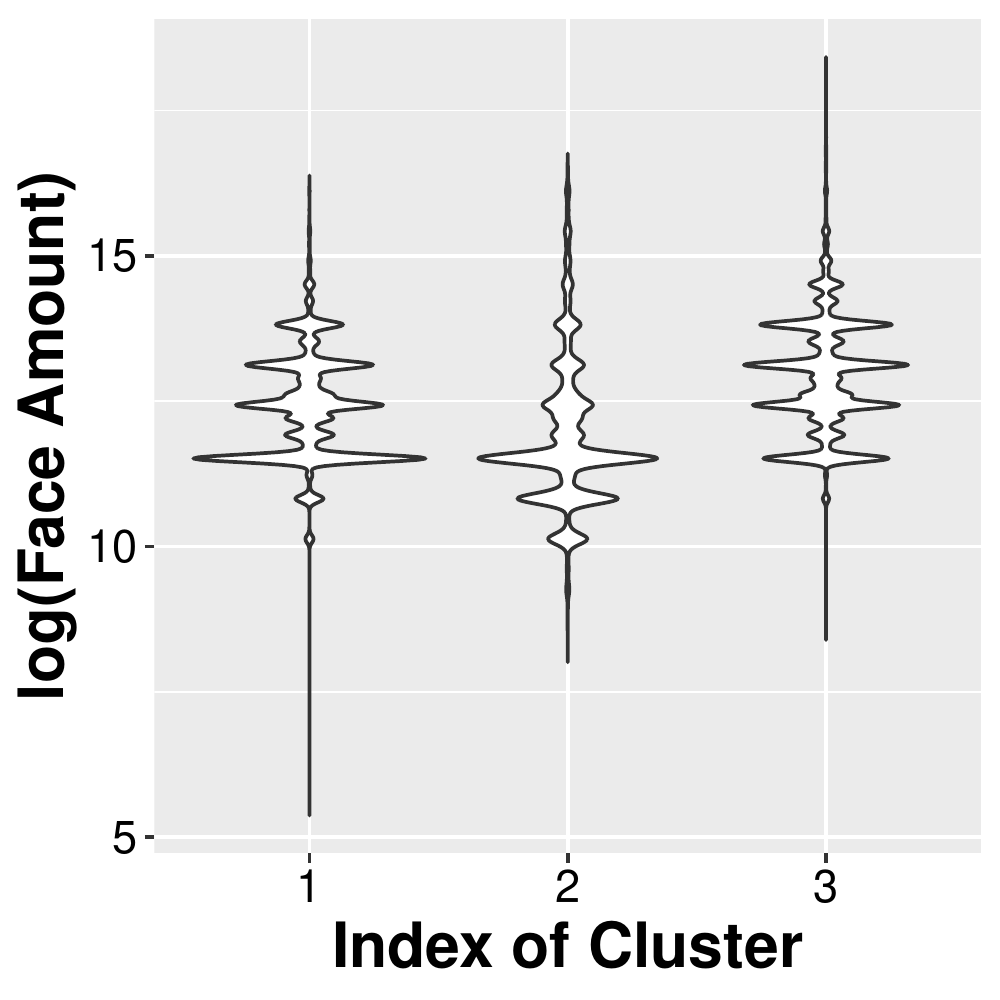}
 \label{fig: faceamount}
\end{subfigure}%
\caption{Distribution of the numerical and categorical attributes in each of the optimal clusters}
\label{fig:variable_cluster}
\end{figure}

Let us describe some dominating features for each of the clusters. The outputs are visualized in Figure \ref{fig:cluster_states} and Figure \ref{fig:variable_cluster}. Additional details of these dominating features are well summarized in Tables \ref{tab:categorical cluster} showing the cluster distribution in the categorical variables, Table \ref{tab:cluster_state} with a descending order of the cluster proportion in the variable States, and Table \ref{tab:numerical cluster} regarding the distributions of numerical variables. These tables are provided in the Appendix.

\begin{enumerate}
\item[] \textbf{Cluster 1}
\begin{itemize}
\item  Its gender make-up is predominantly females in the entire portfolio. There is a larger percentage of Term plan and fewer percentage of Substandard policies than Clusters 2 and 3. The violin plots for the numerical attributes show that the youngest group with smallest amount of insurance coverage is in this cluster. Geographically, the insureds in this cluster are mostly distributed in the northeastern region such as New Jersey, New York, Rhode Island, and New Hampshire.  
\end{itemize}
\item[] \textbf{Cluster 2}
\begin{itemize}
\item This cluster has a gender make-up that is interesting. While Clusters 1 and 3 have a dominating gender, Cluster 2 has 30\% female and 70\% male. It also has the largest proportion of Smokers, Term conversion underwriting type, and Substandard policies when compared with other clusters. However, when it comes to Plan type, 91\% of them have Universal Life contracts and almost no Term plans. With respect to issue age and amount of insurance coverage, this cluster of policies has the most senior people and not surprisingly, it has also lower face amount. Geographically, with exception of few states dominating the cluster, there is almost uniform distribution of the rest of the states in this cluster. Custer 2 have states with the lowest proportion of insured policies among all the clusters.
\end{itemize}
\item[] \textbf{Cluster 3}
\begin{itemize}
\item Male policyholders dominate this cluster and Cluster 3 has the smallest proportion of Smokers and Term Conversion underwriting type among all clusters. More than 90\% of the policyholders purchased Term plan and most of them are also with generally larger face amount than the other two clusters. The policyholders in this cluster are in middle age compared with other clusters according to the violin plots. The policyholders in this cluster are more geographically scattered in Arkansas, Alabama, Mississippi, Tennessee, and Oregon; interestingly, Cluster 3 has the largest proportion of policies among all clusters. 
\end{itemize}
\end{enumerate}

\bigskip

\bigskip

\newpage

\subsection{Analysis of mortality deviation}

We now compare these clusters with respect to their deviations of actual to expected mortality. It is a typical practice in the life insurance industry that when analyzing and understanding such deviations, we compare the actual-to-expected (A/E) death experiences. 

To illustrate how we made the comparison, consider one particular cluster containing $n$ policies. We computed the actual number of deaths for this entire cluster by adding up all the face amounts of those who died during the quarter. Let $\FA_i$ be the face amount of policyholder $i$ in this particular cluster. Thus, the aggregated actual face amount among those who died is equal to
\[
\sum_{i=1}^{n} \text{A}_i = \sum_{i=1}^{n} \FA_i \times I_i,
\]
where $I_i = 1$ indicates the policyholder died and the aggregated expected face amount is
\[
\sum_{i=1}^{n} \E_i = \sum_{i=1}^{n} \FA_i \times q_i,
\]
where the expected mortality rate, $q_i$, is based on the latest 2015 Valuation Basic Table (VBT), using smoker-distinct and ALB (age-last-birthday)\footnote{\url{https://www.soa.org/resources/experience-studies/2015/2015-valuation-basic-tables/}}. The measure of deviation, $R$, is then defined to be
\[
R = \frac{\sum_{i=1}^{n} \text{A}_i}{\sum_{i=1}^{n} \E_i}.
\]
Clearly, a ratio $R <1$ indicates better than expected mortality while $R >1$ indicates worse than expected mortality.

The death indicator $I_i$ is a Bernoulli distributed random variable with parameter $q_i$ which represents the probability of death, or loosely speaking, the mortality rate. For large $n$, i.e., as $n \rightarrow \infty$, the ratio $R$ converges in distribution to a normal random variable with mean $1$ and variance $\frac{\sum_{i=1}^{n} \text{FA}^2_i q_{i}(1-q_{i})}{(\sum_{i=1}^{n} \E_i)^2}$. The details of proofs for this convergence are provided in the appendix.

\begin{figure}[htbp]
    \centering
    \begin{subfigure}{0.8\textwidth}
      \includegraphics[width=\linewidth]{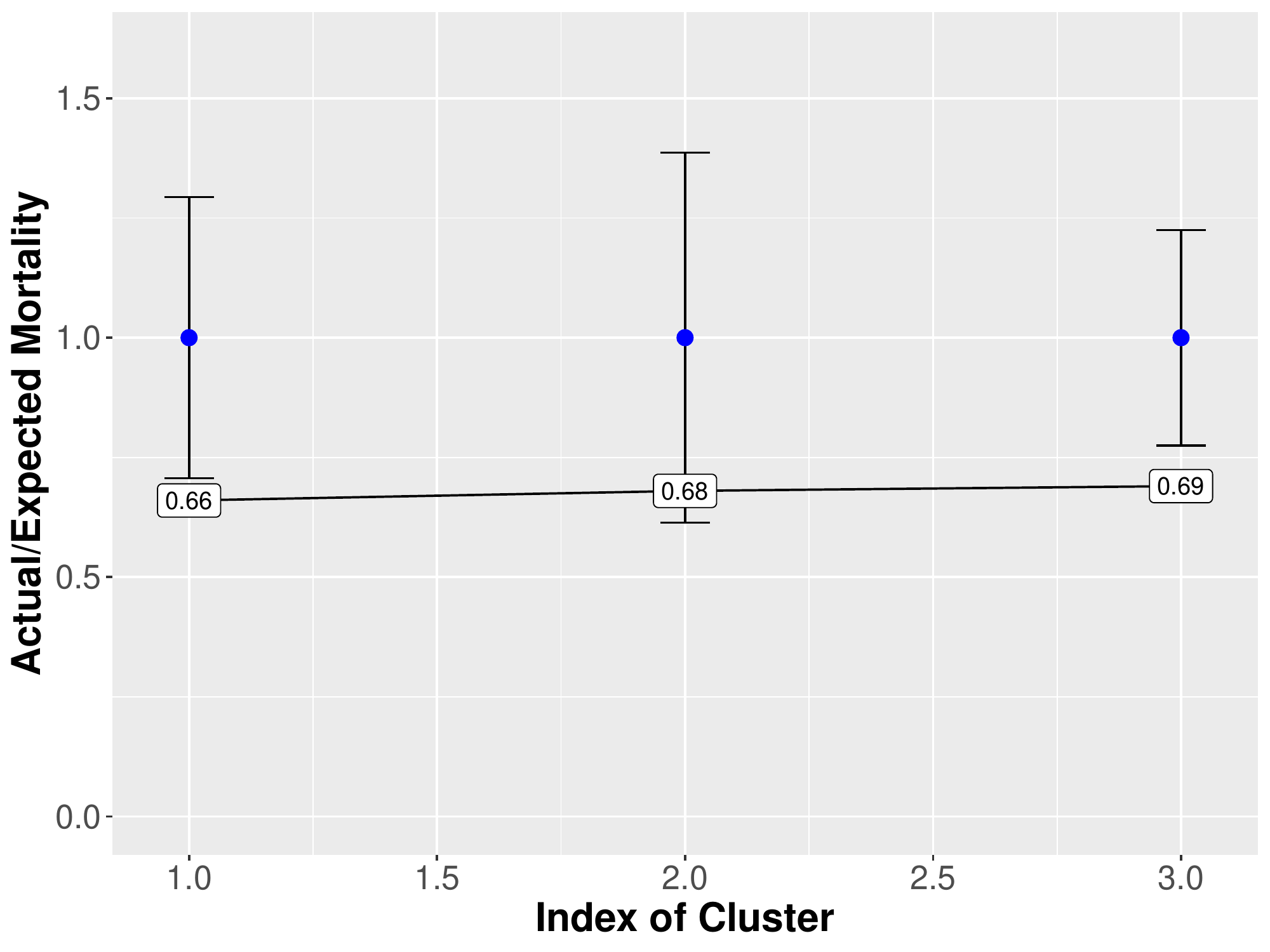}
      \caption{90\% Confidence Interval of A/E Ratio}
      \label{fig:90CI}
    \end{subfigure}
    \begin{subfigure}{0.8\textwidth}
      \includegraphics[width=\linewidth]{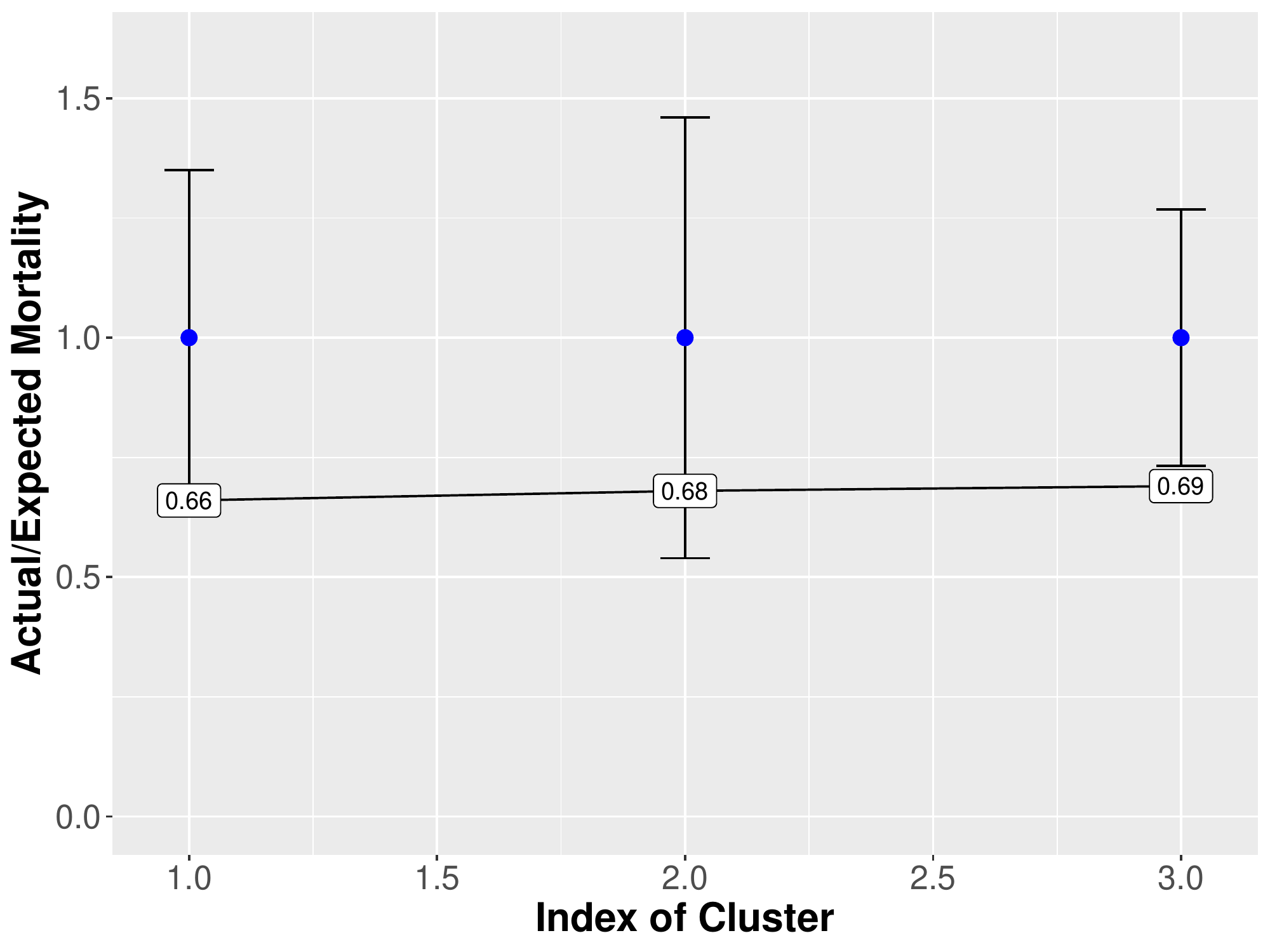}
      \caption{95\% Confidence Interval of A/E Ratio}
      \label{fig:95CI}
    \end{subfigure}
   \caption{Actual to expected mortality rates based on face amounts}
 \label{fig:AECI}
\end{figure}

Based on the results of this convergence, it allowed us to construct 90\% and 95\% confidence interval of the ratio $R$ or the A/E of mortality. We display Figure \ref{fig:90CI} and Figure \ref{fig:95CI}, which depict the differences in the A/E ratio for the three different clusters, based respectively on a 90\% and 95\% confidence intervals, respectively.

Based on this company's claims experience, these figures provide some good news overall. The observed A/E ratios for all clusters are all below 1, which as earlier said indicates that the actual mortality is better than expected for all 3 clusters. There are some peculiar observations that we can draw from the clusters:

\begin{itemize}
\item Cluster 1 has the most favorable A/E ratio among all the cluster and is significantly less than 1 at 10\% significance level, with moderate variability. This can be explained reasonably by this dominant feature compared with other clusters: its gender make-up of all females in the entire portfolio. Female tends to live longer than male on the average. There is a larger percentage of Variable Life plan, slightly fewer Smokers, Term conversion, and Substandard policies than Clusters 2 and 3. In addition, the violin plots for the numerical attributes show that the youngest group with smallest amount of insurance coverage belong to this cluster. We expect this youngest group to have generally low mortality rates. Geographically, the insureds in this cluster are mostly distributed in the northeastern region such as New Jersey, New York, Rhode Island, and New Hampshire.  It may be noted that policyholders tend to come from this region where the population has typically better income with better employer-provided health insurance.

\item Cluster 2 has the A/E ratio of 0.68 and is not significantly less than 1 at both 5\% and 10\% significance levels; it has the largest variability of this ratio among all clusters. Cluster 2 has therefore the most unfavorable A/E ratio from a statistical perspective. The characteristics of this cluster can be captured by these dominant features: (i) Its gender make-up is a mixed of male and female, with more males than females; (ii) It has the largest proportion of Smokers, Term conversion underwriting type and Substandard policies compared with other clusters; (iii) However, when it comes to plan type, 91\% of them have Universal Life contracts and no Term policies; (iv) With respect to issue age and amount of insurance coverage, this cluster has the largest proportion of elder people and therefore, has lower face amounts. All these dominating features help explain a generally worse mortality and larger variability of deviations. For example, the older group has a higher mortality rate than the younger group and with the largest proportion of smokers, this explains the compounded mortality. To some extent, with the largest proportion of Term conversion underwriting types and Substandard policies, they reasonably indicate more inferior mortality experience.

\item Cluster 3 has the A/E ratio most significantly less than 1, even though it has the worst A/E ratio among all the clusters. The characteristics can be captured by some dominating features in the cluster: male policyholders dominate this cluster and it has the smallest proportion of Smokers and Term Conversion underwriting type among three clusters. More than 90\% of the policyholders purchased Term plan and most of them have larger face amount than other clusters. The policyholders in this cluster are about the middle age groups compared to other clusters according to the violin plots. The policyholders are more geographically scattered in Arkansas, Alabama, Mississippi, Tennessee, and Oregon. We generally know that smokers mortality is worse than non smokers. Relatively younger age groups have a lower mortality rate than other age groups. Term plans generally have fixed terms and are more subject to frequent underwriting. The small variability can be explained by having more policies giving enough information and hence, with much more predictable mortality. \end{itemize}

\newpage

\section{Conclusions} \label{sec:conclude}

In this paper, we investigated the use of $k$-prototypes clustering algorithm to provide insights as to the death claims experience of a portfolio of contracts from a life insurance company. Developing a tracking and monitoring system of death claims is an important part of managing a portfolio of life insurance policies. We explore how the results from the $k$-prototypes clustering algorithm can help us detect peculiar characteristics of our life insurance portfolio in order to have an improved understanding of mortality deviations. The $k$-prototypes algorithm integrates the procedures of $k$-means and $k$-modes to efficiently cluster our data set that contains numerical, categorical, and spatial attributes. Our data set consists of a life insurance company's death claims experience observed during the third quarter of 2014, with approximately 1.14 million unique policies and a total insured amount of over 650 billion dollars. The optimal number of clusters are obtained using gap statistics; the algorithm produced three dominating natural clusters in this insurance portfolio. We then used the clusters to compare and monitor actual to expected death claims experience.  Each cluster has a lower actual to expected death claims but with differing variabilities, and each optimal cluster showed patterns of mortality deviation for which we are able to deduce the dominant characteristics of the policies within a cluster. We also find that the additional information drawn from the spatial nature of the policies contributed to an explanation of the deviation of mortality experience from expected. The results can help facilitate decision making because of an improved understanding of potential favorable and unfavorable clusters.

\bigskip

\section*{Acknowledgments}

We thank the financial support of the Society of Actuaries through its Centers of Actuarial Excellence (CAE) grant for our research project on \textit{Applying Data Mining Techniques in Actuarial Science}. We also express our gratitude to Professor Dipak Dey who provided guidance, especially to Shuang Yin, in the completion of this work. He is a faculty member of the Department of Statistics at our university.

\bigskip

\appendix
\numberwithin{equation}{section}
\renewcommand{\theequation}{A.\arabic{equation}}
\section*{Appendix A. Convergence of A/E ratio}

Define  $S_n = X_1 + \cdots X_n$ and $B^2_n  = \Vari(S_n) = \sum_{k=1}^{n} \sigma^2_k $ and for $\epsilon > 0$ let
\[
    L_n(\epsilon) = \frac{1}{B^2_n} \sum_{k=1}^{n} \E (X_k - \mu_k)^2 1_{|X_k - \mu_k|>\epsilon B_n}.
\]
\textbf{Lindeberg-Feller Theorem}: \ Let $\{X_n\}_{n\geq1}$ be a sequence of independent random variable with mean $\mu_n$ and variances $0 < \sigma^2_n <\infty$. If $L_n(\epsilon) \rightarrow 0$ for any $\epsilon >0$, then \\
\begin{equation*}
    \frac{S_n - \E [S_n]}{B_n} \xrightarrow{d} \mathcal{N}(0,1).
\end{equation*}
\textbf{Lyapunov Theorem}: Assume that $\E|X_k|^{2+\delta} < \infty$ for some $\delta >0$ and $k=1,2, \dots.$ If 
\begin{align*}
    \frac{1}{B^{2+\delta}_n}\sum_{k=1}^n \E |X_k-\mu_k|^{2+\delta} \rightarrow 0
\end{align*}
Then
\begin{align*}
    \frac{S_n -\E [S_n]}{B_n} \xrightarrow{d} \mathcal{N}(0,1)
\end{align*} 
\begin{proof}
For $\delta >0$, 
\begin{align*}
    L_n (\epsilon) & = \frac{1}{B^2_n} \sum_{k=1}^n \E(X_k-\mu_k)^2 1_{|X_k-\mu_k|>\epsilon B_n}  = \frac{1}{B^2_n}\sum_{k=1}^n \sum_{k=1}^n \E \frac{|X_k-\mu_k|^{2+\delta}}{|X_k-\mu_k|^{\delta}}1_{|X_k-\mu_k|>\epsilon B_n} \\
    &\leq \frac{1}{\epsilon^{\delta}B^{2+\delta}_n} \sum_{k=1}^n \E|X_k-\mu_k|^{2+\delta} \rightarrow 0 \
    \ \ \text{as} \ \ n\rightarrow \infty
\end{align*}
Then by Lindeberg-Feller Theorem, $\frac{S_n -\E [S_n]}{B_n} \xrightarrow{d} \mathcal{N}(0,1)$. \\
\end{proof}
We can prove that if $\{X_n\}_{n \geq 1}$ is a sequence of independent random variables such that $0<\inf_n \Vari (X_n)$ and $\sup_n \E|X_n|^3 < \infty$. Then $(S_n -\E [S_n])/B_n \xrightarrow{d} \mathcal{N}(0,1)$. \\
\begin{proof}:
Suppose that $X_n$ has mean $\mu_n$ and variance $\sigma_n^2 < \infty$. 
\begin{align*}
   \frac{\sum_{k=1}^n \E |X_k|^3}{B^3_n} = \frac{\sum_{k=1}^n \E |X_k|^3}{(\sum_{k=1}^n \sigma^2_k)^{\frac{3}{2}}} \leq \frac{n \cdot \sup_n \E |X_n|^3}{(n\cdot \inf_n \Vari (X_n))^{\frac{3}{2}}} = \frac{\sup_n \E |X_n|^3}{(\inf_n \Vari (X_n))^{\frac{3}{2}}} \frac{1}{\sqrt{n}} \rightarrow 0 \ \  \text{as $n \rightarrow \infty$}.
\end{align*} 
where $\sup_n \E |X_n|^3 < \infty$ and $0< \inf_n \Vari(X_n) <\infty$. Therefore by Lyapunov Theorem, $(S_n -\E [S_n])/B_n \xrightarrow{d} \mathcal{N}(0,1)$. \\
\end{proof}
In this paper, each policy has a death indicator $I_i$ that is Bernoulli distributed with probability of death $q_{x_i}$. Assume that each policy's death is observable and fixed, not random, so that $q_{x_i}$ is fixed and not varying with data. Within cluster $c$ with total number of policies $n_c$, let $\FA_i$, $\text{A}_i$, and $\E_i$ be the face amount, actual death payment, and expected death payment for each policy, respectively. When a policy $i$ is observed dead, then $I_i=1$. Otherwise, $I_i = 0$. Thus, $\text{A}_i = \FA_i \cdot I_i$ and $\E_i = \FA_i \cdot q_{x_i}$. Let $Y_i =c_i I_i$ where $c_i =\frac{\FA_i}{\sum_{k=1}^{n_c} \E_k}$. Since $I_i \sim \text{Bernoulli}(q_{x_i})$, $\E(Y_i) = c_i\E(I_i)=c_i q_{x_i}$ and $\Vari (Y_i) = c^2_i q_{x_i}(1-q_{x_i})$. We calculate that $\inf_n \Vari(X_n) = 1.67*10^{-16}$, then $\inf_n \Vari(Y_n)$ is positive and finite, and $0 < \sup_n \E|Y_n|^3 = 1.2*10^{-5} < \infty$. These two conditions are satisfied and $Y_i$'s are independently distributed.

Let $R_c = \sum_{i=1}^{n_c} Y_i=\frac{\sum_{i=1}^{n_c} A_i}{\sum_{i=1}^{n_c} \E_i}$ denote the measure of mortality deviation for cluster $c$. By Lyapunov Theorem, we have
\begin{equation*}
\frac{\sum_{i=1}^{n_c} Y_i -  \E(\sum_{i=1}^{n_c} Y_i)}{\sqrt{ \Vari(\sum_{i=1}^{n_c} Y_i)}} \xrightarrow{d} \mathcal{N}(0,1) \Rightarrow R_c \xrightarrow{d} \mathcal{N} \left (1, \frac{\sum_{i=1}^{n_c} \text{FA}^2_i q_{x_i}(1-q_{x_i})}{(\sum_{i=1}^{n_c} \E_i)^2}\right), 
\end{equation*}
where
\begin{equation*}
\E(R_c)=\sum_{i=1}^{n_c} \E(Y_i)=\sum_{i=1}^{n_c} c_i q_{x_i}= \frac{\sum_{i=1}^{n_c} \text{FA}_i * q_{x_i}}{\sum_{i=1}^{n_c} \E_i}= \frac{\sum_{i=1}^{n_c} \E_i}{\sum_{i=1}^{n_c} \E_i}=1
\end{equation*}
and
\begin{equation*}
\Vari(R_c) = \sum_{i=1}^{n_c}  \Vari (Y_i) = \sum_{i=1}^{n_c} c^2_i q_{x_i}(1-q_{x_i})=\frac{\sum_{i=1}^{n_c} \text{FA}^2_i q_{x_i}(1-q_{x_i})}{(\sum_{i=1}^{n_c} E_i)^2}.
\end{equation*}

\newpage

\section*{Appendix B. Tables that summarize the distribution of clusters}

\input{clust_state}

\begin{table}[htbp]
  \centering
  \caption{Data summary for categorical variables within the 3 optimal clusters}
  \scalebox{0.82}{
    \begin{tabular}{rl|ccccc}
    \toprule
    \toprule
          & \multicolumn{1}{l}{\textbf{Categorical Variables}} & \textbf{Levels } & \textbf{Cluster 1} & \textbf{Cluster 2} & \textbf{Cluster 3} \\
    \midrule
          & \multicolumn{1}{l|}{\textbf{Gender}} & Female & 100\% & 30.43\% & 0.09\% \\
          &       & Male  & 0\% & 69.57\% & 99.91\%  \\
          &       &       &       &       &       &  \\
          & \multirow{1}[0]{*}{\textbf{Smoker Status}} & Smoker & 4.43\% & 6.53\% & 3.45\%  \\
          &       & Nonsmoker & 95.57\% & 93.47\% & 96.55\% \\
          &       &       &       &       &       &  \\
          & \multirow{1}[0]{*}{\textbf{Underwriting Type }} & Term conversion & 3.59\% & 22.79\% & 0.85\%\\
          &       & Underwritten & 96.41\% & 77.21\% & 99.15\% \\
          &       &       &       &       &       &  \\
          & \multirow{1}[0]{*}{\textbf{Substandard Indicator }} & Yes & 6\% & 11.58\% & 7.82\% \\
          &       & No    & 94\% & 88.42\% & 92.18\%  \\
          &       &       &       &       &       &  \\
          & \multirow{1}[1]{*}{\textbf{Plan}} & Term  & 73.71\% & 0\% & 91.51\% \\
          &       & ULS   & 8.95\% & 90.86\% & 0.12\% \\
          &       & VLS   & 17.34\% & 9.14\% & 8.37\%  \\
    \bottomrule
    \bottomrule
    \end{tabular}%
    }
  \label{tab:categorical cluster}%
\end{table}%

\begin{table}[htbp]
  \centering
  \caption{Data summary for numerical variables within the 3 optimal clusters}
  \scalebox{0.75}{
    \begin{tabular}{lp{9.785em}ccccc}
    \toprule
    \toprule
    \textbf{Continuous Variables} & \multicolumn{1}{c}{} & \textbf{Minimum} &\textbf{1st Quantile} & \textbf{Mean} & \textbf{3rd Quantile} & \textbf{Maximum } \\
    \midrule
    \multirow{4}[1]{*}{\textbf{Issue Age}} & \textbf{Cluster 1} & 0  & 31 & 38.59 & 46  & 81 \\
          & \textbf{Cluster 2} & 0  &  47 & 51.53 & 61 & 90 \\
          & \textbf{Cluster 3} & 0    & 36 & 44.47 & 53 & 85 \\
          & \multicolumn{1}{r}{} &       &       &  \\
    \multirow{4}[1]{*}{\textbf{Face Amount}} & \textbf{Cluster 1} & 215 & 100,000 &375,066 & 500,000 & 13,000,000 \\
          & \textbf{Cluster 2} & 3,000 & 57,000  & 448,634 &  250,000 & 19,000,000\\
          & \textbf{Cluster 3} & 4,397 & 250,000 &  717, 646 & 1,000,000 & 100,000,000 \\
    \bottomrule
    \bottomrule
    \end{tabular}%
    }
  \label{tab:numerical cluster}%
\end{table}%


\newpage

\bibliographystyle{apalike}
\bibliography{dcclust}

\end{document}

%% file: clust_state.tex
\begin{table}[htbp]
  \centering
  \caption{Proportions of each cluster in the variable \textit{States}}
  \scalebox{0.62}{
    \begin{tabular}{cc|cc|cc}
    \multicolumn{2}{c|}{Cluster 1 } & \multicolumn{2}{c|}{Cluster 2} & \multicolumn{2}{c}{Cluster 3} \\
    States & proportion & States & proportion & States & proportion \\
    \midrule
    NJ    & 36.36\% & WV    & 21.25\% & AR    & 74.78\% \\
    NY    & 34.54\% & DE    & 19.40\% & AL    & 74.19\% \\
    RI    & 34.35\% & PA    & 19.26\% & MS    & 73.84\% \\
    NH    & 34.09\% & OH    & 18.50\% & TN    & 71.36\% \\
    ME    & 33.98\% & IN    & 16.40\% & OR    & 70.64\% \\
    MA    & 33.64\% & RI    & 16.21\% & ID    & 69.03\% \\
    DE    & 32.70\% & ME    & 15.79\% & OK    & 68.16\% \\
    CA    & 32.46\% & SD    & 15.48\% & KY    & 68.02\% \\
    NV    & 32.25\% & IL    & 15.45\% & TX    & 66.90\% \\
    MD    & 31.90\% & NJ    & 14.17\% & WA    & 66.29\% \\
    IL    & 31.82\% & NY    & 14.12\% & UT    & 64.42\% \\
    PA    & 31.63\% & SC    & 13.84\% & GA    & 64.34\% \\
    DC    & 31.54\% & MD    & 13.54\% & CO    & 64.29\% \\
    CT    & 30.82\% & IA    & 12.78\% & WY    & 63.83\% \\
    MT    & 29.96\% & MO    & 12.78\% & ND    & 63.20\% \\
    NM    & 29.94\% & KS    & 12.69\% & NE    & 62.92\% \\
    IN    & 29.55\% & ND    & 12.51\% & NC    & 62.69\% \\
    FL    & 29.19\% & LA    & 12.32\% & KS    & 61.93\% \\
    MN    & 28.60\% & VT    & 12.22\% & VA    & 61.80\% \\
    AZ    & 28.47\% & MA    & 12.15\% & IA    & 61.74\% \\
    WI    & 28.34\% & FL    & 12.07\% & LA    & 61.48\% \\
    MI    & 27.94\% & NH    & 11.93\% & MT    & 61.02\% \\
    VT    & 27.93\% & MN    & 11.65\% & MO    & 60.85\% \\
    OH    & 27.61\% & WI    & 11.65\% & AZ    & 60.69\% \\
    WY    & 27.12\% & MI    & 11.64\% & MI    & 60.42\% \\
    UT    & 27.01\% & NM    & 11.59\% & WI    & 60.01\% \\
    VA    & 26.91\% & CT    & 11.53\% & SD    & 59.86\% \\
    SC    & 26.86\% & NE    & 11.51\% & VT    & 59.85\% \\
    WA    & 26.61\% & NC    & 11.49\% & MN    & 59.75\% \\
    MO    & 26.37\% & VA    & 11.29\% & SC    & 59.31\% \\
    LA    & 26.20\% & CA    & 11.26\% & FL    & 58.74\% \\
    GA    & 26.10\% & AZ    & 10.83\% & DC    & 58.66\% \\
    NC    & 25.82\% & NV    & 10.34\% & NM    & 58.47\% \\
    CO    & 25.66\% & CO    & 10.05\% & CT    & 57.65\% \\
    NE    & 25.57\% & KY    & 9.98\% & NV    & 57.40\% \\
    IA    & 25.48\% & DC    & 9.80\% & CA    & 56.28\% \\
    WV    & 25.48\% & GA    & 9.56\% & MD    & 54.56\% \\
    KS    & 25.38\% & OK    & 9.08\% & MA    & 54.21\% \\
    SD    & 24.66\% & WY    & 9.05\% & IN    & 54.05\% \\
    ND    & 24.29\% & MT    & 9.02\% & NH    & 53.98\% \\
    TX    & 24.24\% & TX    & 8.86\% & OH    & 53.90\% \\
    ID    & 22.84\% & AL    & 8.74\% & WV    & 53.27\% \\
    OK    & 22.77\% & MS    & 8.62\% & IL    & 52.73\% \\
    OR    & 22.24\% & UT    & 8.57\% & NY    & 51.34\% \\
    KY    & 22\%  & ID    & 8.13\% & ME    & 50.23\% \\
    TN    & 20.71\% & TN    & 7.93\% & NJ    & 49.47\% \\
    AR    & 18.04\% & AR    & 7.18\% & RI    & 49.44\% \\
    MS    & 17.54\% & OR    & 7.12\% & PA    & 49.10\% \\
    AL    & 17.07\% & WA    & 7.10\% & DE    & 47.90\% \\
    \bottomrule
    \end{tabular}%
    }
  \label{tab:cluster_state}%
\end{table}%